\documentclass[twocolumn,twocolappendix]{aastex63}

\received{}
\revised{}
\accepted{}
\submitjournal{ApJ}

\shorttitle{Gas dynamics in the star forming region G18.148$-$0.283}
\shortauthors{Dey et al.}

\graphicspath{{./}{figures/}}

\begin{document}

\title{Gas dynamics in the star forming region G18.148$-$0.283: Is it a manifestation of two colliding molecular clouds?}

\correspondingauthor{Jyotirmoy Dey}
\email{jyotirmoydey.18@res.iist.ac.in}


\author{Jyotirmoy Dey}
\affiliation{Department of Earth and Space Sciences, Indian Institute of Space Science and Technology, Trivandrum, Kerala 695547, INDIA}

\author{Jagadheep D. Pandian}
\affiliation{Department of Earth and Space Sciences, Indian Institute of Space Science and Technology, Trivandrum, Kerala 695547, INDIA}

\author{Dharam Vir Lal}
\affiliation{National Centre for Radio Astrophysics - Tata Institute of Fundamental Research, Post Box 3, Ganeshkhind P.O., Pune 411007, INDIA}

\begin{abstract}

We report the results obtained from a multi-wavelength study of the HII region, G18.148$-$0.283, using the \textit{upgraded Giant Metre-wave Radio Telescope} (uGMRT) at 1350 MHz along with other archival data. In addition to the radio continuum emission, we have detected the H169$\alpha$ and H170$\alpha$ radio recombination lines towards G18.148$-$0.283 using a correlator bandwidth of 100 MHz. The moment-1 map of the ionized gas reveals a velocity gradient of approximately 10~km~s$^{-1}$ across the radio continuum peaks. The $^{12}$CO ($J$=3$-$2) molecular line data from the COHRS survey also shows the presence of two velocity components that are very close to the velocities detected in the ionized gas. The spectrum and position-velocity diagram from CO emission reveal molecular gas at an intermediate velocity range bridging the velocity components. We see mid-infrared absorption and far-infrared emission establishing the presence of a filamentary infrared dark cloud, the extent of which includes the targeted HII region. The magnetic field inferred from dust polarization is perpendicular to the filament within the HII region. We have also identified two O9 stars and 30 young stellar objects towards the target using data from the 2MASS, UKIDSS, and GLIMPSE surveys. Cumulatively, this suggests that the region is the site of a cloud-cloud collision that has triggered massive star formation and subsequent formation of an HII region.

\end{abstract}


\section{Introduction} 
\label{sec:section1}

The understanding of the formation processes of massive stars ($M \geq $ 8$-$10$\, M_\odot$) or star clusters is still far from complete. Observational studies targeting massive star formation are more challenging compared to their low-mass counterparts due to massive stars forming on short time scales and in clustered environments \citep{2006MNRAS.373.1563K}. Since the massive star forming regions are on average more distant compared to the areas forming low-mass stars, there are additional limitations of angular resolution and sensitivity of observing facilities \citep{1994ApJS...91..659K,2002ARA&A..40...27C}. Despite these difficulties, studies of the massive star formation are crucial, as they continuously influence the evolution of galaxies by emitting protostellar jets, ejecting stellar materials, accreting gas, and various other mechanisms.

Although the formation of low or intermediate-mass stars by accretion of matter is reasonably well understood, one cannot simply scale up this process to form massive stars. The significant number of companions around high-mass stars influence the subsequent accretion onto the massive protostellar cores \citep{2009Sci...323..754K}. The outward radiation pressure can also halt the accretion process before the core reaches its final mass. Thus, the massive core needs to accrete mass with a high accretion rate ($10^{-4}$--$10^{-3}$~$M_\odot$~yr$^{-1}$; \citealt{1987ApJ...319..850W, 2003ApJ...585..850M}) to overcome the radiation pressure.

\citet{2003ApJ...585..850M} explained that such an accretion rate is possible if the surface density of the clumps ($\Sigma$) is sufficiently high with a value $\sim$ 1~g~cm$^{-2}$. One of the ways to create a region with such high values of $\Sigma$ is a cloud-cloud collision (CCC) event. In this process, two supersonic clouds collide into one other to create a shocked and compressed region suitable for fragmentation and further collapse. CCCs are not very rare in the Milky Way. Early observations of CCC include studies of NGC 133, where the CO spectra revealed two velocity components \citep{1976ApJ...209..466L, 1977ApJ...215..129L}. The first cluster to be identified as a possible collision object was Westerland 2, which harbors two associated clouds with a velocity difference of $20$~km~s$^{-1}$ \citep{2009ApJ...696L.115F, 2010ApJ...709..975O} approximately. Other recent observations by \cite{2014ApJ...780...36F}, \cite{2015ApJ...806....7T}, \cite{2016ApJ...833...85B}, \cite{2017ApJ...849...65D}, \cite{2018ApJ...861...19D, 2019ApJ...878...26D}, \cite{2020MNRAS.499.3620I} have also detected CCCs in the high-mass star forming regions.

In most cases, the relative velocity between the two colliding clouds leads to two different velocity peaks or components in the resulting molecular spectrum \citep{1976ApJ...209..466L, 1977ApJ...215..129L,1978ApJ...223..840D}. Since the mixing of the colliding clouds continues for a long time after the instance of the collision, a collision-front of shocked materials having intermediate velocities also forms in between the colliding clouds. As a result of this event, a ``broad-bridge'' feature appears in a position-velocity (PV) diagram of the colliding region \citep{2015MNRAS.450...10H} that connects the velocity peaks detected in the molecular spectrum. Thus, surveys of molecular line emission (especially $^{12}$CO) are great tools to identify and verify CCC events. One such survey is the \textit{CO High-Resolution Survey} (COHRS; \citealt{2013ApJS..209....8D}), which maps the $^{12}$CO ($J$=3$-$2) transition in the Galactic plane between $-0.5\degr$ $\leq$ \textit{b} $\leq 0.5\degr$ and 10$\degr$ $\leq$ \textit{l} $\leq$ 65$\degr$ with a velocity resolution of 1~km~s$^{-1}$.

Based on their magnetohydrodynamic simulations, \citet{2013ApJ...774L..31I} also concluded that a collision between two supersonic clouds gives rise to the dense filamentary structures enhancing the surface density and self-gravity inside the colliding clouds. Such a filament may achieve the required high surface density leading to the formation of massive stars. Simulations also show that the collision amplifies the magnetic field in a direction perpendicular to that of the filament. Although the perpendicular alignment of magnetic fields is found to be general outcome of turbulent magnetohydrodynamic simulations (e.g. \citealt{2019MNRAS.485.4509L} and references therein), detection of such a field geometry using the dust polarization data from the \textit{Planck} mission\footnote{\url{https://www.cosmos.esa.int/web/planck}} can also act as a subsidiary signature of a CCC along with other pieces of evidence.  

Massive stars also ionize their surrounding environment by producing enough UV photons, creating HII regions. Observations of the HII regions using radio recombination lines (RRL) help us determine the ionized gas's kinematics and properties, such as electron temperature. Moreover, in the event of a CCC, one may observe two velocity components in the RRL emission if massive stars form from both clouds around the collision interface.

In this paper, we perform a multi-wavelength study of a HII region, G18.148$-$0.283 (G18.15 henceforth), to understand its gas dynamics and formation. Located at a heliocentric distance of 4.1~kpc \citep{2006ApJ...653.1226Q}, G18.15 has a physical diameter of 4.9~pc, which corresponds to an angular diameter of 4.1$\arcmin$ for the given distance. G18.15 is located in the first Galactic quadrant ($l$ $=$ $18.148\degr$, $b$ $=$ $-0.283\degr$; $\alpha$ $=$ $18^{\text{h}}25^{\text{m}}01.01^{\text{s}}$, $\delta$ $=$ $-13\degr15\arcmin33.5\arcsec$). The \textit{Co-Ordinated Radio 'N' Infrared Survey for High-mass star formation} (CORNISH; \citealt{2012PASP..124..939H}; \citealt{2013ApJS..205....1P}) reported a flux density of $856.18\pm82.85$~mJy at 5~GHz, and classified G18.15 as an ultracompact-HII (UC-HII) region. \citet{2006ApJ...653.1226Q} used the National Radio Astronomy Observatory (NRAO) 140-ft (43~m) telescope in Green Bank at 8.6~GHz (HPBW = $3.20\arcmin$) to estimate an electron temperature ($T_{\text{e}}$) of $7180\pm70$~K for the entire region assuming G18.15 to be a homogeneous, isothermal sphere. \citet{1989ApJS...71..469L} measured the local standard of rest (LSR) velocity of G18.15 using the same 140-ft telescope, and reported $V_{\text{LSR}}$ $=$ $53.9\pm0.4$~km~s$^{-1}$. \citet{1994ApJS...91..659K} estimated the lower limit of the Lyman-continuum photon rate in the region based on which they assigned a spectral type of B0.5 to the central ionizing star.

In their multi-wavelength-based observation, \citet{2017RAA....17...57Z} concluded that there are three clump candidates associated with G18.15, and high-mass star formation processes may still be happening within these clumps. According to them, the formation process was triggered by the forward-propagating shock wave, which is originated from the expansion of G18.15 itself. In this paper, we present evidence that the star formation activity may be the outcome of a CCC.

The structure of our paper is as follows. In section~\ref{sec:section3}, we describe our radio observations with the uGMRT and the ancillary data used for our study. In section~\ref{sec:section4}, we describe the results of our study, including radio continuum, RRL emission, distribution of molecular hydrogen, ionizing stars, and young stellar objects (YSO) in the region. Lastly, in section~\ref{disc}, we discuss the possibility of a CCC event behind the formation of G18.15.



\section{Observations and Archival Data} \label{sec:section3}

\subsection{uGMRT observations}

Our observation of G18.15 was carried out using the \textit{upgraded Giant Metrewave Radio Telescope} (uGMRT) \citep{1990IJRSP..19..493S} situated at Pune, India. The observations were carried out with the GWB correlator configured to have a bandwidth of 100~MHz centered at 1350~MHz with 8192 channels. uGMRT has a native resolution of 2$\arcsec$ and a largest detectable angular scale of 7$\arcmin$ in this band. The details of the observational run are furnished in Table~\ref{tab:table1}. A total of five hydrogen RRLs, H167$\alpha$ to H171$\alpha$ (see Table 2) are present in this frequency range.
The radio sources 3C48 was used as the flux density calibrator and bandpass calibrator, and 1911$-$201 was used as the gain calibrator.
 
\begin{deluxetable}{lc}
\tablenum{1}
\label{tab:table1}
\tablecaption{Details of our observation using uGMRT.}
\tablewidth{0pt}
\tablehead{
\colhead{Parameter} & \colhead{Value}
}
\startdata
        Target name & G18.148-0.283 \\
		Observation date & March 11, 2019\\
		System Temperature & 73 K \\
		On-source time & 210 minutes\\
		No. of channels & 8192\\
		Central frequency & 1350 MHz\\
		Bandwidth & 100 MHz \\
		Primary Beam  & $25\arcmin$ \\
		Synthesised Beam & $\approx2\arcsec$ \\
		Peak continuum flux density &  373.3 mJy beam$^{-1}$ \\
		Theoretical rms ($\sigma$)$^{\text{a}}$ & 0.4 mJy beam$^{-1}$ \\
\enddata
\tablecomments{a = in a single channel at 10 km s$^{-1}$ resolution before stacking.}
\end{deluxetable}
 
The data were reduced using the NRAO \textit{Common Astronomy Software Applications} (CASA 4.1; \citealt{2007ASPC..376..127M}) package. An initial bandpass solution was computed and applied to the data in order to identify and excise RFI. The bad data were then flagged successively using the FLAGDATA task.
Next, gain solutions were computed and applied to the calibrators after which another iteration of data examination and editing was carried out. The final bandpass and gain solutions were computed using this data, after which the solutions were applied to the target source.

Although the RFI affected channels were flagged from the target data before calibration, another careful examination of the data was performed before imaging. After RFI excision, the net bandwidth available for imaging G18.15 is around 50~MHz considering only the line-free channels. The target was then imaged using the CLEAN task, and the continuum image was self-calibrated to improve the dynamic range. The final 1$\sigma$ noise in the continuum image is 0.06~mJy~beam$^{-1}$.
 
As shown in Table 1, the observations were carried out with 8192 channels, providing a native velocity resolution $\sim2.6$~km~s$^{-1}$. The gain calibrated line data were first Hanning-smoothed to a velocity resolution $\sim5$~km~s$^{-1}$ in order to get rid of the Gibbs ringing phenomenon caused by the strong RFI sources. Next, we used the UVLSF task of the NRAO \textit{Astronomical Image Processing System} (AIPS) package to subtract the continuum from the total emission, as UVLSF gave a better result than the UVCONTSUB task of CASA. The self-calibration solutions determined from the radio continuum were then applied to the line data.
 
After this, the H169$\alpha$ and H170$\alpha$ line data were imaged using CLEAN. The H167$\alpha$ and H171$\alpha$ could not be imaged due to the lines being affected by the edges of the bandpass. The H168$\alpha$ line was discarded due to it having poor signal to noise ratio. To enhance the signal to noise ratio, the images of both RRLs were stacked to the observed frequency of H169$\alpha$ using the package Line-Stacker \citep{2020MNRAS.499.3992J}\footnote{\url{https://github.com/jbjolly/LineStacker/releases}}. This results in an improvement in the signal to noise ratio of the line image by roughly a factor of $\sqrt{2}$. The final rms noise (= $\sigma$) after stacking is equal to 0.25~mJy~beam$^{-1}$.

\subsection{Archival Data}

The uGMRT observations are complemented by data from \textit{The HI/OH/Recombination line survey of the inner Milky Way} (THOR; \citealt{2016A&A...595A..32B}). The THOR survey was carried out with the \textit{Karl G. Jansky Very Large Array} (VLA) in the C-configuration using the WIDAR correlator covering the HI 21-cm line, four OH, and nineteen RRLs in addition to the continuum emission at 1--2~GHz in full polarization.
The detected RRLs are also stacked in order to increase the signal to noise ratio. The final spectral line map from THOR has a velocity resolution of 10~km~s$^{-1}$ and an angular resolution of 40$\arcsec$. There are six different spectral windows to observe the continuum emission at 1--2~GHz. As the angular resolution has a frequency dependence, all continuum images are smoothed to a common resolution of 25$\arcsec$.

We have used $^{12}$CO ($J$=3$-$2) data from the \textit{CO High-Resolution Survey} (COHRS; \citealt{2013ApJS..209....8D}) to study the molecular emission associated with G18.15. The COHRS data were taken using the HARP (\textit{Heterodyne Array Receiver Programme}; 325--375~GHz) receiver on the \textit{James Clerk Maxwell Telescope} (JCMT) in Hawaii. COHRS provides data products at a velocity and angular resolution of 1~km~s$^{-1}$ and 16$\arcsec$ respectively with a typical rms in antenna temperature of $\sim$ 1~K.

To determine the orientation of the magnetic field lines around G18.15, we have utilized the \textit{Planck} 353-GHz (850 $\mu$m) dust continuum polarization data \citep{2015A&A...576A.104P}. The polarization data includes Stokes I, Q, U maps from the \textit{Planck}-Public Data Release 3 full mission maps with PCCS2 Catalog\footnote{\url{https://irsa.ipac.caltech.edu/applications/planck/}}.

To analyze the cold dust emission, we have used data from the Hi-GAL survey \citep{2010PASP..122..314M}. The Hi-GAL survey provides images of the far-infrared continuum at 500, 350, 250, 160 and 70~$\mu$m using the SPIRE and PACS cameras of the Herschel Space Observatory.
We used the level 2.5 data products from the Herschel Science Archive to analyze the dust emission and estimate the hydrogen column density ($N(\text{H}_2)$) across G18.15.



We have also used the mid-infrared point source catalog from the GLIMPSE survey \citep{2003PASP..115..953B}, and near-infrared point source catalogs from the 2MASS \citep{2005sptz.prop20710S} and \textit{UKIDSS Galactic Plane Survey} \citep{2007MNRAS.379.1599L} to study the YSO population and main sequence stars in G18.15.


\begin{figure}
   \includegraphics[width=\columnwidth]{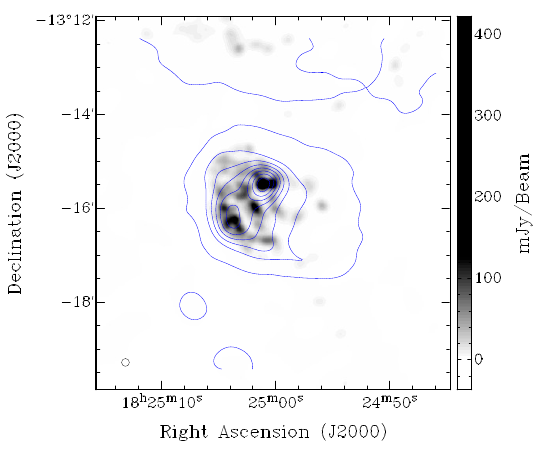}
   \caption{ The radio continuum image of G18.15 at 1350~MHz is shown overlaid with the contour levels (in blue) from THOR at 1310~MHz. The color scale depicts the specific intensity or brightness in $\mathrm{mJy\,beam^{-1}}$. The levels are starting from 3$\sigma$ and increasing in steps of 75~$\mathrm{mJy\,beam^{-1}}$, while the rms noise is equal to $\sim$~1.4 $\mathrm{mJy\,beam^{-1}}$. The respective beam size in black empty circle is shown at the bottom-left corner of the figure.}
   \label{fig:figure1}
\end{figure}

\begin{deluxetable}{lc}
\tablenum{2}
\label{tab:table2}
\tablecaption{Name and rest frequencies of the RRLs that were targeted in our observation.}
\tablewidth{0pt}
\tablehead{
\colhead{RRL Name} & \colhead{Rest frequency ($\nu_0$ in MHz)}
}
\startdata
    	H167$\alpha$ & 1399.368 \\
		H168$\alpha$ & 1374.601 \\
		H169$\alpha$ & 1350.414 \\
		H170$\alpha$ & 1326.792 \\
		H171$\alpha$ & 1303.718 \\
\enddata
\end{deluxetable}

\begin{figure*}
   \epsscale{1}
   \plotone{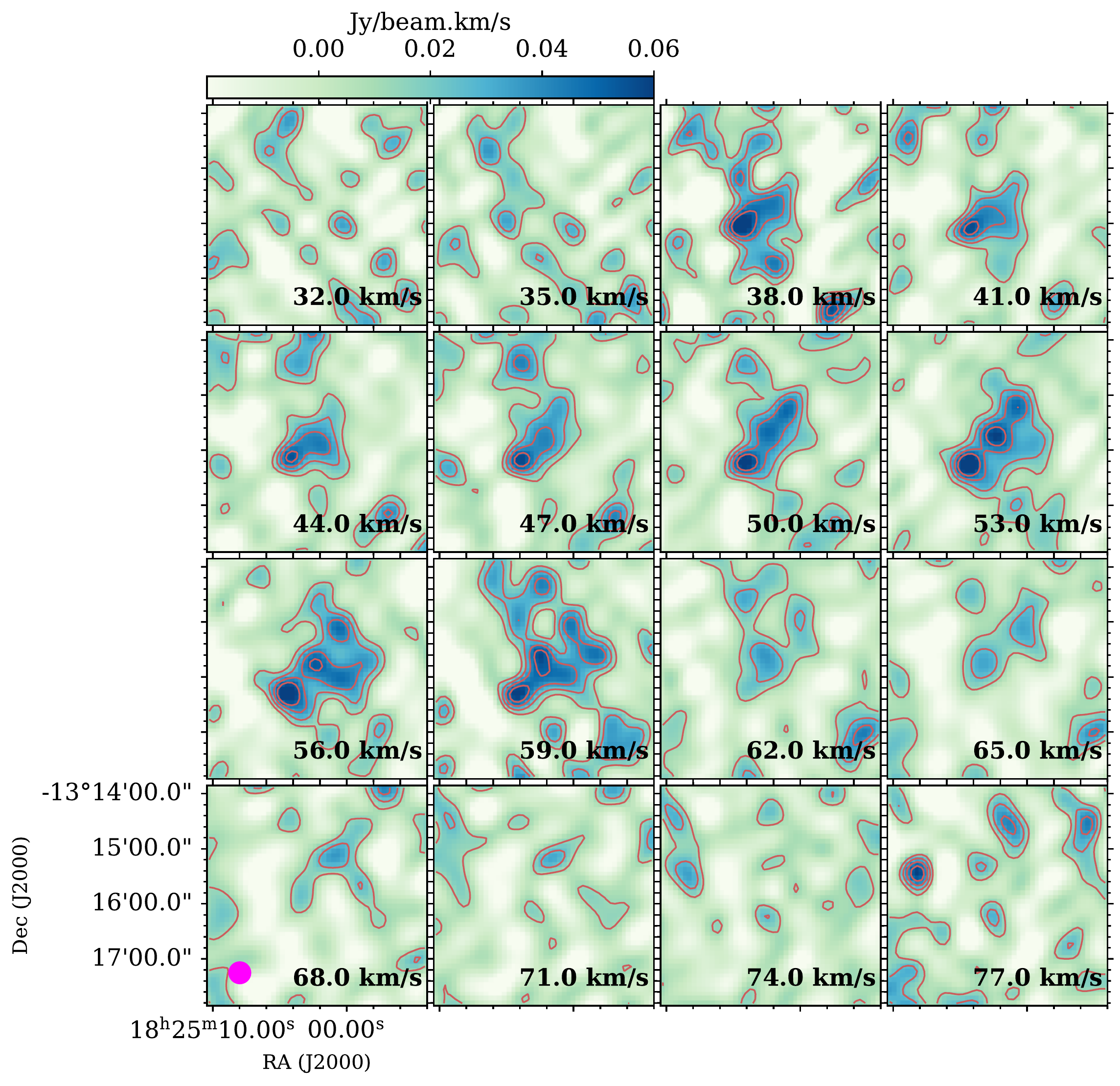}
   \caption{The stacked RRL channel map towards G18.15 is shown here. Each panel shows the velocity integrated intensity within a velocity range of 3 km s$^{-1}$. The 25$\arcsec$ beam is shown in the filled magenta circle.}
   \label{fig:figure3}
\end{figure*}

\begin{figure}
   \includegraphics[width=\columnwidth]{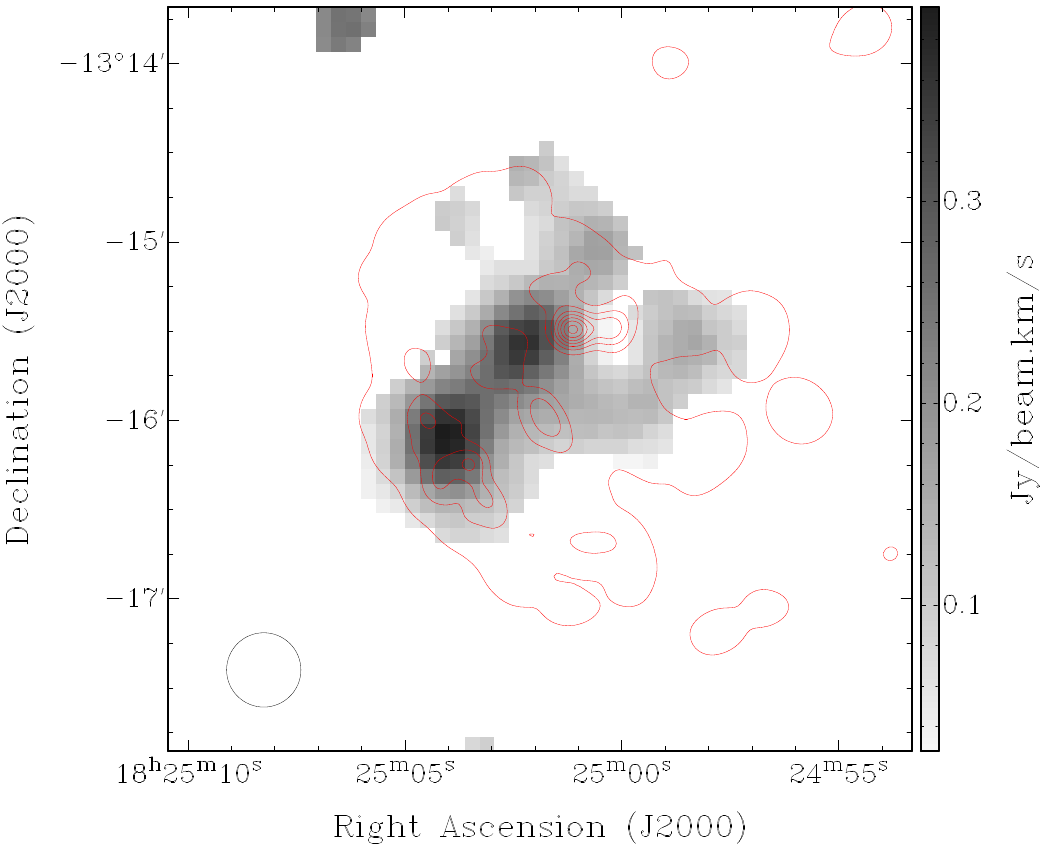}
   \caption{The moment-0 map from the uGMRT RRL data is shown overlaid with the respective radio contours in red from the continuum emission. The contours are increasing in steps of 50 $\mathrm{mJy\,beam^{-1}}$ starting from the 3$\sigma$ level. The restoring beam is shown in black empty circle at the bottom-left corner of the figure.}
   \label{fig:moment0}
\end{figure}

\section{Results}
\label{sec:section4}
\subsection{Radio emission from the ionized gas}

\subsubsection{Continuum emission}
\label{sec:subsubsec1}
Figure~\ref{fig:figure1} shows the 1350~MHz radio continuum emission from G18.15 at 10$''$ resolution using the uGMRT, with the THOR map at 25$''$ resolution being overlaid in contours. Although the native resolution of uGMRT is 2$''$, we have restricted the resolution to 10$''$ to recover the extended emission. The morphology of radio continuum emission is elongated along the NW-SE direction with an extended envelope of emission along the perpendicular direction of the central elongated structure. At 25$''$ resolution, there are two peaks in the radio continuum, which are resolved into multiple peaks at the higher resolution of uGMRT.

The spectral index for the region is obtained from the THOR survey, and is seen to be $\approx 0.1$.
This suggests that the emission is thermal, and has an optical depth around unity since one expects the spectral index to be close to 0 as the emission transitions from being optically thick to optically thin.
Although the assumption of the radio continuum emission being optically thin is not strictly valid, we can derive a lower limit for the Lyman-continuum photon rate ($N_{\text{Ly}}$) using the following equation \citep{Mezger67,2016A&A...588A.143S},
\begin{equation}
    N_{\text{Ly}} \geq 4.76 \times 10^{42} \, \nu^{0.1} \, d^2 \, S_\nu \, {T_{\text{e}}}^{-0.45}
    \label{eq:eq1}
\end{equation}
where $S_\nu$ is the flux density in Jansky (Jy), $\nu$ is the frequency in GHz, $d$ is the distance to the source in parsec (pc), and $T_{e}$ is the electron temperature of the ionized gas. Adopting an electron temperature of 7180~K \citep{2006ApJ...653.1226Q} and a distance of 4.1~kpc, one obtains the Lyman continuum photon rate to be larger than $5.8\times10^{48}$~s$^{-1}$ (log~$N_{\text{Ly}}=48.76$). If all the ionizing radiation were to arise from a single star, this would require a main sequence star of spectral type O6.5-O7 or earlier \citep{2005A&A...436.1049M}.

Our estimate of the Lyman continuum photon rate is in good agreement with that of \citet{2017RAA....17...57Z}, who found log~$N_{\text{Ly}}=48.68$ using the radio continuum data from the 1.4~GHz \textit{NRAO VLA Sky Survey} (NVSS; \citealt{1998AJ....115.1693C}), but is significantly higher than that of \citet{1994ApJS...91..659K}, who found log~$N_{\text{Ly}}=46.82$ using observations of the region at 8.4~GHz with the VLA in the B-configuration. A higher Lyman continuum photon rate estimated in our study than that of \citet{1994ApJS...91..659K} is probably due to the extended emission being resolved out in their high-resolution study. Similarly, the Lyman continuum photon rate estimated from the high-resolution ($\sim2\arcsec$) CORNISH survey is seen to be log~$N_{\text{Ly}}=48.0$, which suggests that some extended emission is resolved out in the CORNISH maps as well.


\subsubsection{RRL emission}
\label{sec:subsubsec2}

Figures~\ref{fig:figure3} and~\ref{fig:moment0} show the channel map and integrated intensity map of the RRL emission at an angular resolution of 25$''$. 
It can be seen from the figures that three different emission peaks are aligned with the central elongation of continuum emission.
The SE peak is the brightest among the three peaks, followed by the middle and NW peaks. The SE peak is located within $2\arcsec$ of the continuum SE peak, while the middle and NW peaks are located at opposite sides of the continuum NW peak. 


\subsubsection{Electron temperature map}
\label{sec:subsubsec3}

\begin{figure}
   \includegraphics[width=\columnwidth]{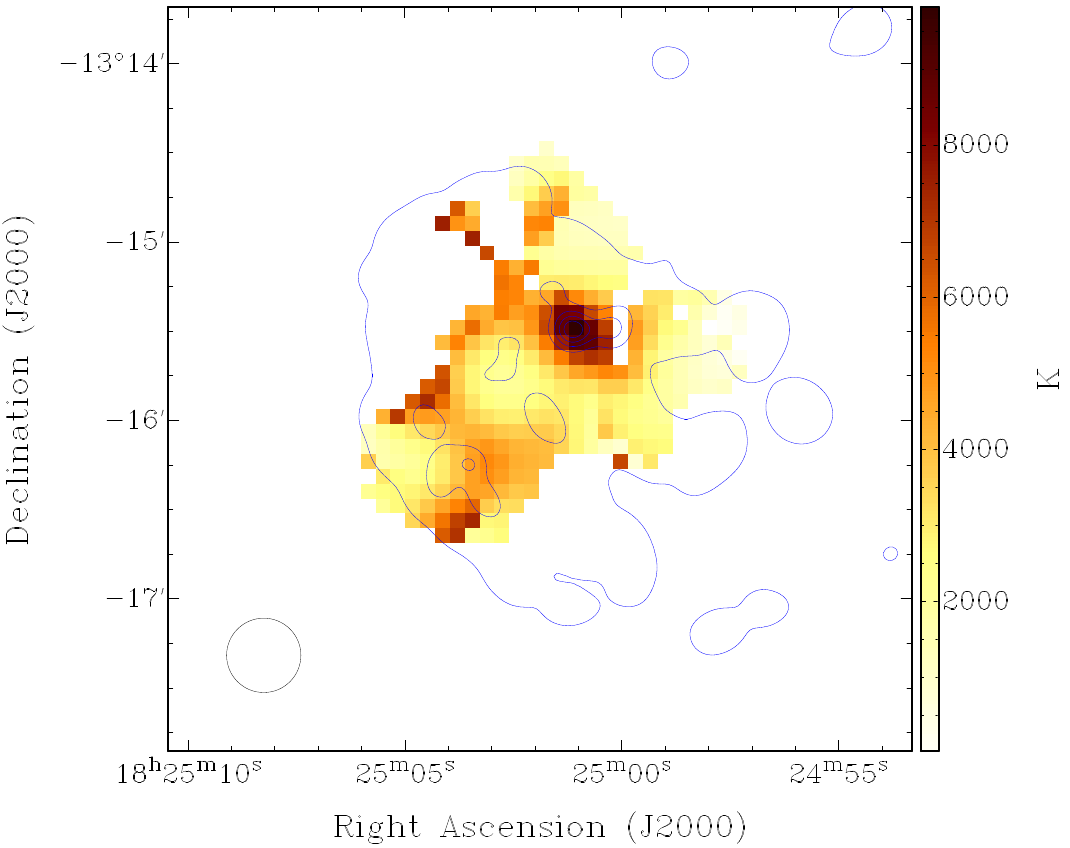}
   \caption{The pixel-wise electron temperature ($T_{\text{e}}$) map (non-LTE) from uGMRT is shown overlaid with the 3$\sigma$ contours (in blue) from the radio continuum emission. The contour levels are increasing in steps of 75 $\mathrm{mJy\,beam^{-1}}$ starting from the 3$\sigma$ level. The beam is shown in black empty circle at the bottom-left corner of the figure.}
   \label{fig:figure6}
\end{figure}

Figure~\ref{fig:figure6} shows the electron temperature ($T_{\text{e}}$) distribution across G18.15 using the RRL and continuum data (the methodology and relevant equations are given in Appendix A). This was done by smoothing the radio continuum map to the same resolution as that of the RRL map -- 25$''$ for the uGMRT data.

The electron temperature at the peaks of continuum emission range from 5200--9500~K. The electron temperature agrees well with the average electron temperature of 7180~K \citep{2006ApJ...653.1226Q} towards the region, considering that this was measured at a much coarser resolution of 3.2$'$. It is, however, surprising to see that the electron temperature is much lower at the locations of diffuse emission.

\subsection{Kinematics of the region}
\label{sec:subsection4.3}
\begin{figure*}
    \includegraphics[width=\columnwidth]{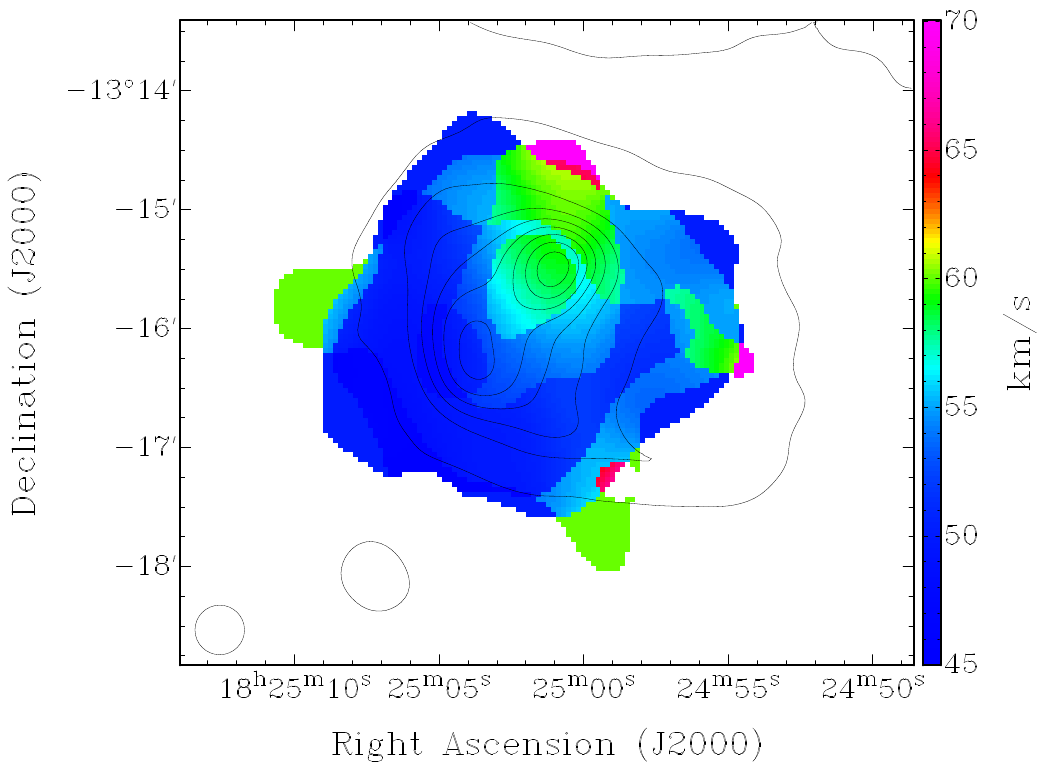}\includegraphics[height=6.3cm,width=8.2cm]{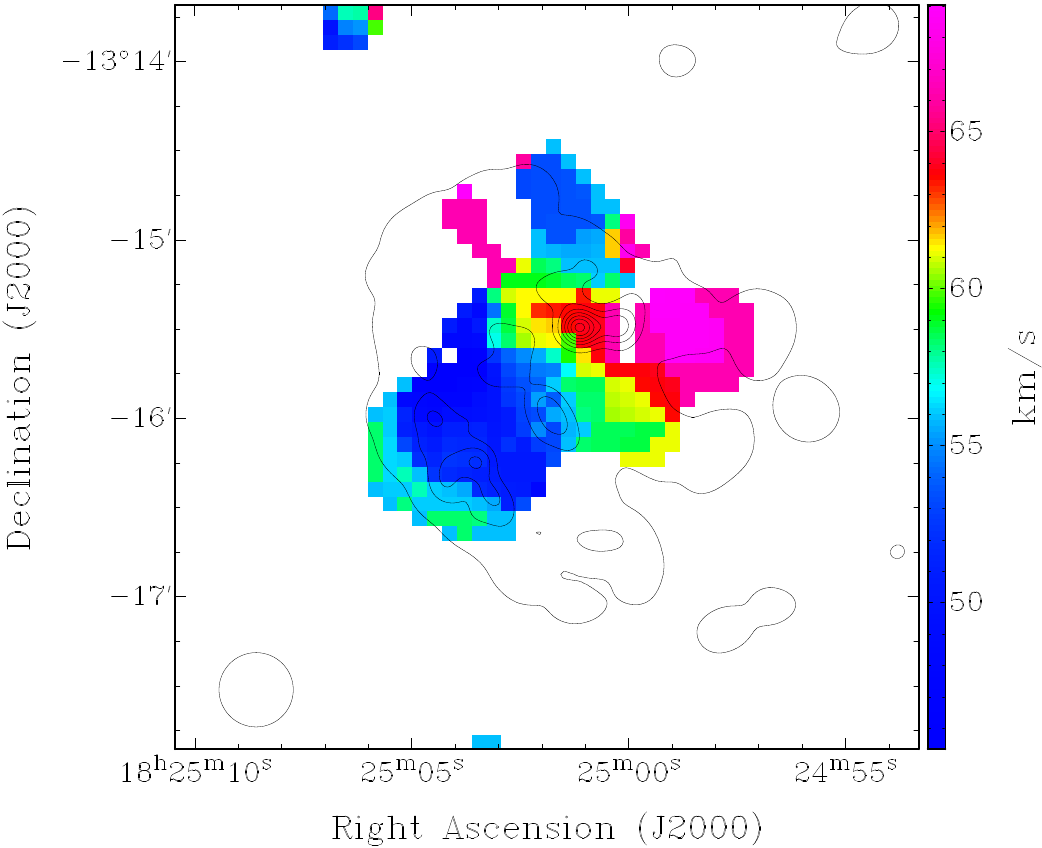}
    \caption{The left and right panels show the moment-1 maps from THOR and uGMRT respectively. A velocity gradient is seen between the NW and SE peaks. The contours in the left and right panel are identical to those in Figure~\ref{fig:figure1} and Figure~\ref{fig:moment0} respectively.}
    \label{fig:figure9}
\end{figure*}

Figure~\ref{fig:figure9} shows the velocity field of the ionized gas in the region. The NW continuum peak is observed to have a higher velocity compared to the SE peak. The velocity field (52--65~km~s$^{-1}$) between the NW and SE peaks appears to be smooth with a linear gradient at 40$''$ resolution of THOR, while at 25$''$ resolution, it appears somewhat discontinuous. The presence of two velocity components to the RRL emission separated by 13~km~s$^{-1}$ with a discontinuity in the middle of the HII region is suggestive of G18.15 being a case of a CCC.

\begin{figure}
    \epsscale{1}
	\plotone{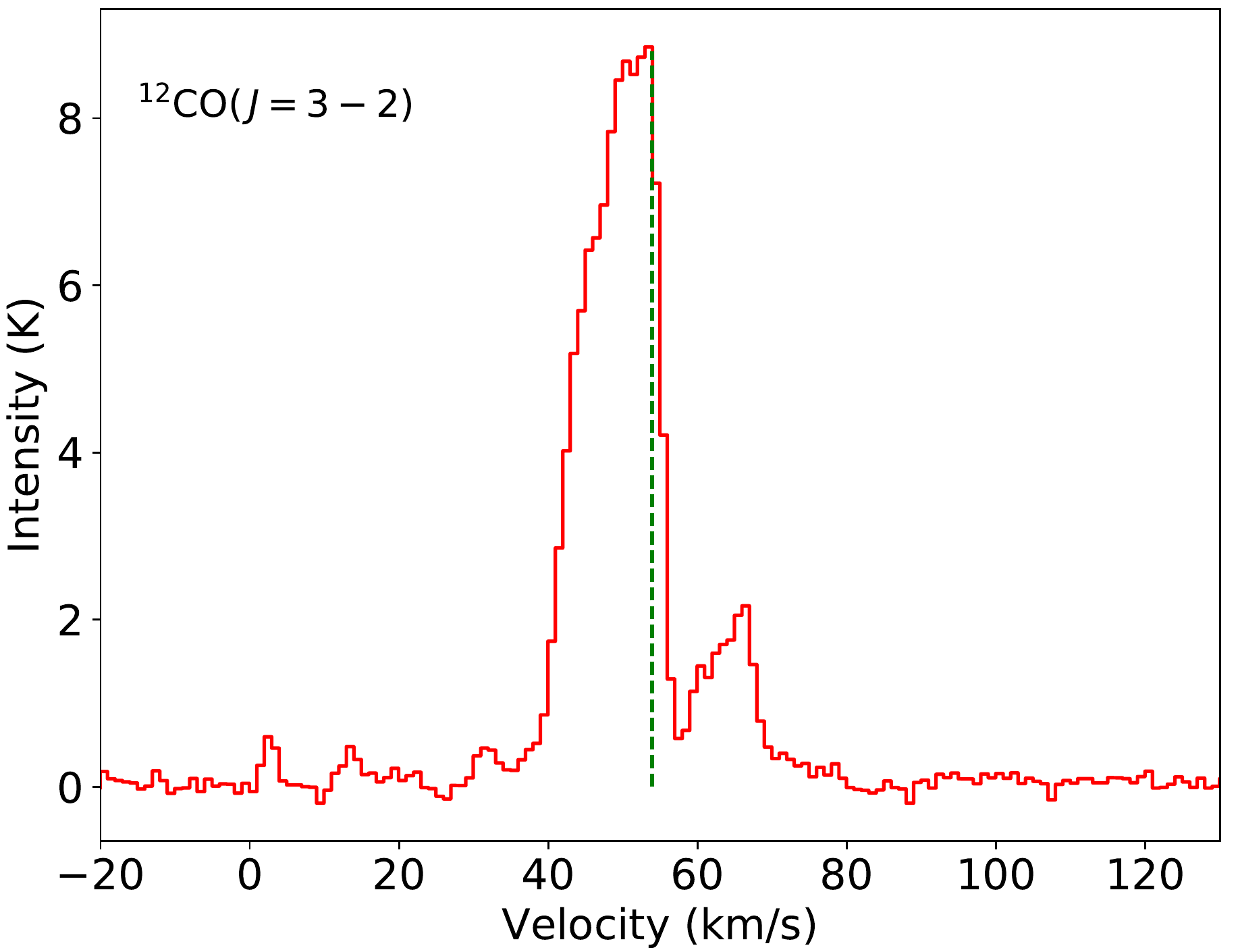}
    \caption{The $^{12}$CO spectrum towards G18.15 shows two velocity peaks at the LSR velocities of 53.4 and 66.7~km~s$^{-1}$ respectively. The intermediate velocities form a narrow plateau connecting those peaks. The LSR velocity measured by \citet{1989ApJS...71..469L} is shown by the green dashed line.}
    \label{fig:figure8}
\end{figure}

\begin{figure*}
	\epsscale{1}
    \plotone{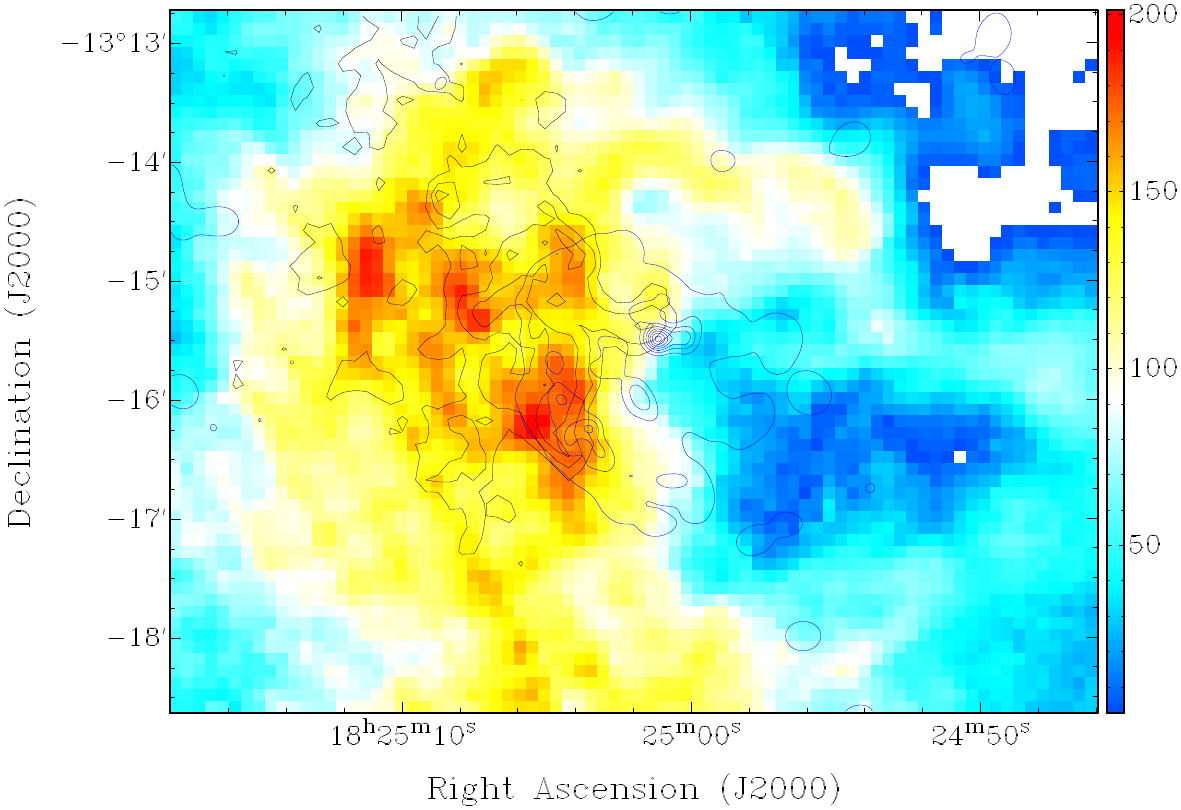}
    \caption{The color scale shows the $^{12}$CO ($J$=3$-$2) integrated intensity $\mathrm{[K\,.\,km\,s^{-1}]}$ within the velocity 39 to 55 km s$^{-1}$. The contours in black (from 18 to 40~K.km~s$^{-1}$ in steps of 8~K.km~s$^{-1}$) show the CO integrated emission within the velocity 59 to 74~km~s$^{-1}$. The contours in blue show the radio continuum emission from uGMRT at 1350~MHz with the contour levels identical to those in Figure~\ref{fig:moment0}.}
    \label{fig:figure2}
\end{figure*}

The possibility of G18.15 being a case of a CCC can be tested by examining the kinematics of the molecular gas in the region. We used the COHRS data for this purpose since the $J$=3$-$2 transition of CO is an excellent tracer of the warm gas (10$-$50~K) at densities around $10^4$~cm$^{-3}$. The CO spectrum towards G18.15 (Figure~\ref{fig:figure8}) shows the presence of two velocity components (or peaks) at 53.4~km~s$^{-1}$ and 66.7~km~s$^{-1}$ respectively. It can be seen that the velocity components in CO are close to those observed in the ionized gas. This suggests that the massive stars in the region have formed from the two molecular clouds traced by the CO emission. The spectrum also shows that the velocity components representing two different molecular clouds are connected by a narrow plateau of intermediate velocity and moderate intensity, which may arise from the interaction between the two clouds. 

The Figure~\ref{fig:figure2} shows the CO integrated intensity map using the velocity range of 39$-$55~km~s$^{-1}$ covering the low-velocity molecular cloud overlaid with the black contours from the integrated intensity map using the 59$-$74~km~s$^{-1}$ velocity range of the high-velocity molecular clouds. The radio continuum emission at 1350~MHz is also overlaid with the blue contours. Figure~\ref{fig:figure2} shows that both clouds are obliquely shaped with a NE-SW elongation.
The figure also shows the presence of a cavity (diameter $\sim$ 0.5~pc) towards the west of the radio continuum peaks with most of the CO line emission occurring from a C-shaped region that is offset to the east of G18.15 from the center.

The presence of two velocity components connected with an intermediate-velocity emission profile in the CO spectrum, a large cavity, a linear gradient and a C-shaped emission region suggest that G18.15 is the site of a CCC event between two molecular clouds with a relative velocity of $10$~km~s$^{-1}$. The features observed are in agreement with the simulations of \citet{1992PASJ...44..203H} and \citet{2018PASJ...70S..58T}. 

\begin{figure}
   \includegraphics[width=\columnwidth]{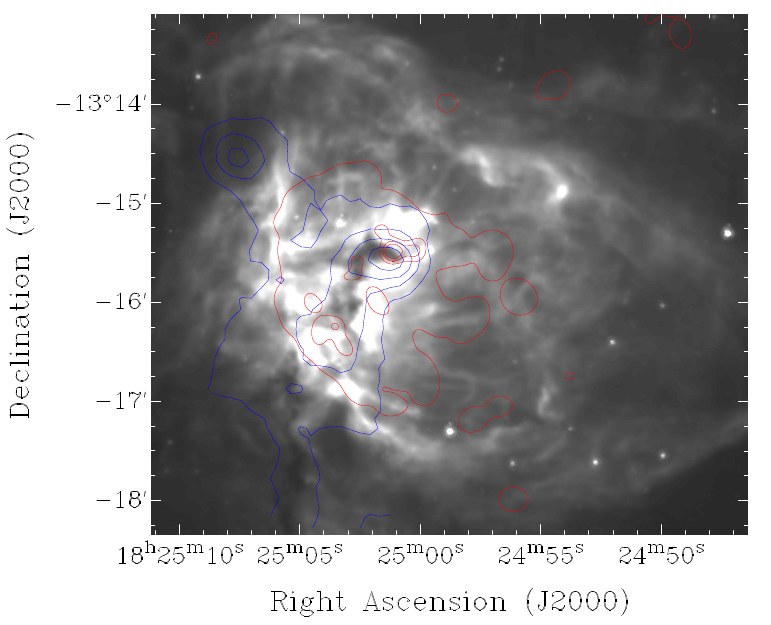}
   \caption{The figure shows the 8.0-$\mu$m mid-infrared image of G18.15 obtained from the \textit{Spitzer}-GLIMPSE survey. The blue and red contours show the dust emission at 870~$\mu$m and radio continuum emission at 1350~MHz respectively.}
   \label{fig:irdc}
\end{figure}

In particular, it is interesting to note the presence of a dark filamentary structure connecting the radio continuum peaks in 8.0~$\mu$m (Figure~\ref{fig:irdc}). Although the filament is not classified as an infrared dark cloud (IRDC) in the catalogs of \citet{2006ApJ...639..227S} and \citet{2009A&A...505..405P}, the extinction at 8.0~$\mu$m that has close correspondence with 870~$\mu$m emission suggests the filament to be an IRDC. This is consistent with the simulations of \citet{2013ApJ...774L..31I} wherein dense filaments are formed in CCC events due to enhancement of the gas surface density.

\begin{figure*}
   \includegraphics[width=\columnwidth]{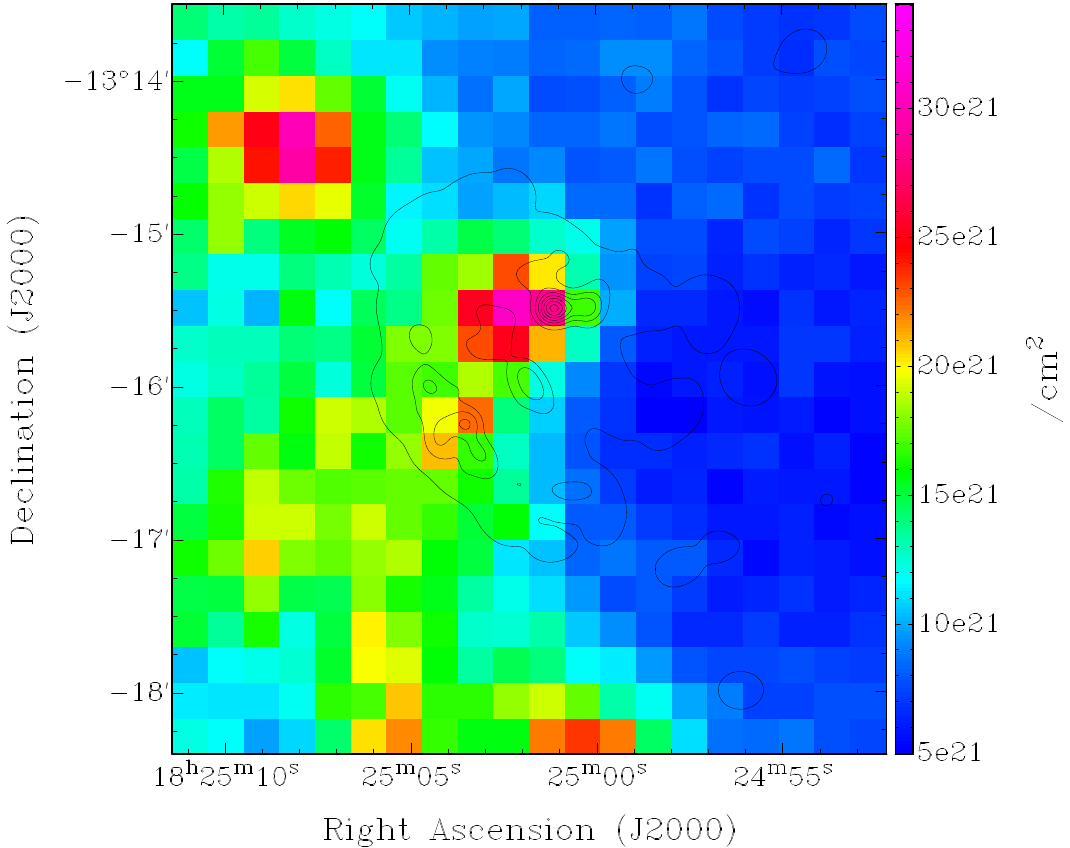}\includegraphics[width=\columnwidth]{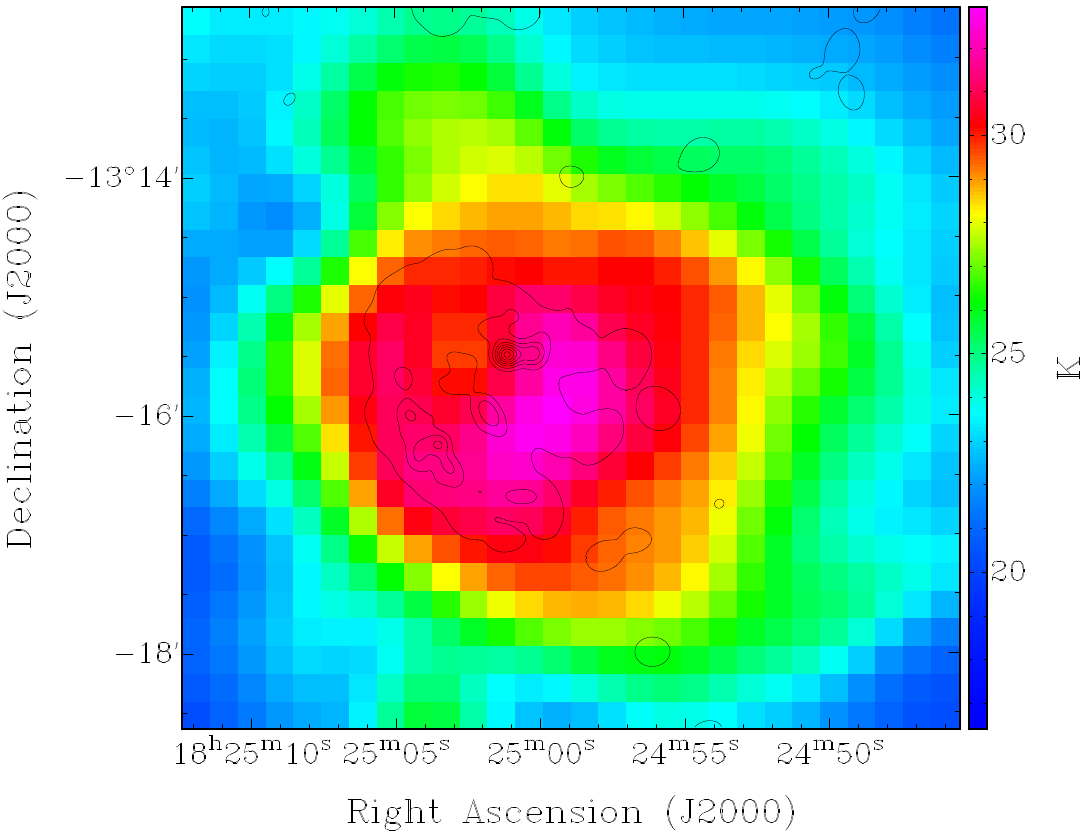}
   \caption{The figure shows the H$_2$ column density (left) and dust temperature (right) maps towards G18.15. The maps are obtained after fitting pixel-wise SEDs to the PACS and SPIRE far-infrared images. The black contours show the radio continuum emission at 1350~MHz starting at the 3$\sigma$ level.}
   \label{fig:figure13}
\end{figure*}

\begin{figure}
	\includegraphics[width=\columnwidth]{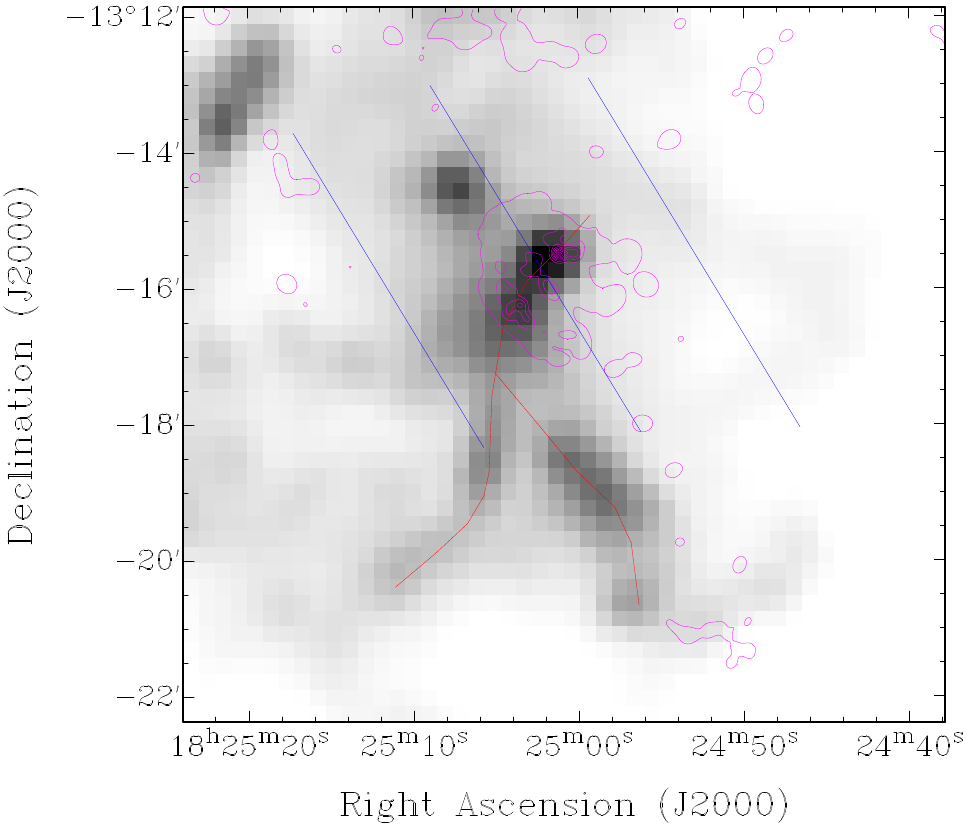}
    \caption{The figure shows the 500-$\mu$m SPIRE image of the cold dust emission overlaid with the 3$\sigma$ radio contours at 1350~MHz in magenta. The red solid lines show visually identified skeleton of the filament, which hosts G18.15. The blue lines are the mean orientation of the magnetic field lines measured around G18.15.}
    \label{fig:figure20}
\end{figure}

\subsection{FIR emission from the cold dust}
\label{sec:fir}

The cold dust emission was analyzed using data from the Hi-GAL survey. We used level 2.5 data from the Herschel Science Archive for this purpose. The data have a plate scale of 3.2$''$ per pixel at the PACS wavelengths of 70 and 160~$\mu$m, and 6$''$, 10$''$, and 14$''$ per pixel at the SPIRE wavelengths of 250, 350 and 500~$\mu$m respectively. Moreover, the data have different units with the PACS data having units of Jy~pixel$^{-1}$, while the SPIRE data having units of MJy~sr$^{-1}$. The data were processed in the Herschel Interactive Processing Environment (HIPE)\footnote{HIPE is a joint development by the \textit{Herschel Science Ground Segment Consortium}, consisting of ESA, the NASA \textit{Herschel Science Center}, and the HIFI, PACS and SPIRE consortia.} to have the same units and were regridded to a common plate scale using a premade kernel \citep{2011PASP..123.1218A}. 
The pixel size of the 500-$\mu$m image (14$\arcsec$) is considered as the reference pixel size, as it has the poorest resolution of 36.4$\arcsec$. A constant background was then subtracted from the data.

To estimate the column density from the far-infrared images, we have fitted a modified black-body function on a pixel-to-pixel basis.
The modified black-body function with specific intensity, $I_\nu$, has the following form,

\begin{equation}
    \label{eq:eq15}
    {I}_\nu = {B}_\nu({T}_{\text{d}})\,\left(1-{e}^{-\tau_\nu}\right)
\end{equation}
where,

\begin{equation}
    \label{eq:eq16}
    \tau_\nu = \frac{\mu\, {m}_{\text{H}}\, \kappa_\nu\, {N}(\text{H}_2)}{\eta}
\end{equation}
Here $\tau_\nu$ is the optical depth of the cold medium at frequency $\nu$, ${B}_\nu({T}_{\text{d}})$ is the black-body function at the dust temperature $T_{\text{d}}$, $\eta$ is the gas-to-dust ratio of the ISM, which is $\sim 100$, $\mu$ is the mean weight of the molecular gas, $\kappa_\nu$ is the dust opacity, $m_{\text{H}}$ is the mass of a hydrogen atom, and $N$(H$_2$) is the column density of molecular hydrogen. We have taken $\mu$ to be 2.86 assuming that the cold molecular gas is made up with 70\% molecular hydrogen by mass \citep{2010A&A...518L..92W}. Following \citet{1994A&A...291..943O}, the dust opacity at frequency $\nu$ can be expressed by the following equation,

\begin{equation}
    \label{eq:eq17}
    \kappa_\nu = \kappa_0\, \left(\frac{\nu}{\nu_0}\right)^\beta
\end{equation}
where $\beta$ is the dust emissivity index. We have assumed $\beta=2$ following \citet{2013A&A...554A..42R}, and $\kappa_0=5.04$~cm$^{2}$~g$^{-1}$ at 500~$\mu$m for a MRN-distribution of grain sizes (for the diffuse interstellar medium of gas density $=10^6$~cm$^{-3}$) with thin ice mantles \citep{1994A&A...291..943O}. We have considered $T_{\text{d}}$ and $N(\text{H}_2$) as free parameters while fitting the far-infrared data points using the non-linear least squares Marquardt-Levenberg algorithm. The column density and dust temperature maps are shown in the left and right panels of Figure~\ref{fig:figure13} respectively.

High column densities can be observed across the radio continuum peaks tracing the central filament with a peak value of $3.1\times10^{22}$~cm$^{-2}$ located close to the NW continuum peak. The dust temperature also peaks towards the center and remains almost constant overall ($\sim 33$~K) within the HII region. The relatively high value of dust temperature suggests significant heating from the massive stars inside the HII region.

\begin{deluxetable*}{llllcccccc}
\tablenum{3}
\label{tab:table4}
\tablecaption{List of candidate ionizing stars detected within G18.15}
\tablewidth{0pt}
\tablehead{
\colhead{Name} & \colhead{Designation} & \colhead{$\alpha_{\text{J2000}}$} & \colhead{$\delta_{\text{J2000}}$} &
\colhead{J} & \colhead{H} & \colhead{K} & \colhead{Spectral type} & \colhead{$N$(H$_2$)$^{\text{a}}$} & \colhead{$N$(H$_2$)$^{\text{b}}$}\\
\colhead{} & \colhead{} & \colhead{(deg)} & \colhead{(deg)} &
\colhead{(mag)} & \colhead{(mag)} & \colhead{(mag)} & \colhead{} & \colhead{$(\times10^{22}$cm$^{-2}$)} & \colhead{$(\times10^{22}$cm$^{-2})$}
}
\startdata
2MASS-1 & 18250164-1315414 & 276.257 & -13.2615 & 12.253 & 10.664 & 9.76 & B2 & 2.81 & 2.12\\
2MASS-2 & 18250397-1316152 & 276.267 & -13.2709 & 14.207 & 13.66 & 13.326 & O9 & 1.28 & 1.96 \\
2MASS-3 & 18250091-1316532 & 276.254 & -13.2815 & 12.759 & 12.18 & 11.857 & B3 & 1.16& 1.18\\
2MASS-4 & 18250457-1315191 & 276.269 & -13.2553 & 14.689 & 13.745 & 13.187 & B3 & 1.63 & 1.75\\
2MASS-5 & 18250585-1315478 & 276.274 & -13.2633 & 14.597 & 13.827 & 13.436 & B2 & 1.56 & 1.51\\
UKIDSS-1 & 438754351144 & 276.273 & -13.262 & 15.666 & 14.885 & 14.430 & O9 & 1.63 & 1.66\\
UKIDSS-2 & 438754352028 & 276.279 & -13.256 & 14.523 & 13.890 & 13.518 & B3 & 1.27 & 1.25\\
\enddata
\tablecomments{\, a $=$ estimated from eq.~\ref{eq:equ19a}, b $=$ estimated from the respective locations in Figure~\ref{fig:figure17a}.}
\end{deluxetable*}

   \begin{figure*}
   \includegraphics[width=\columnwidth]{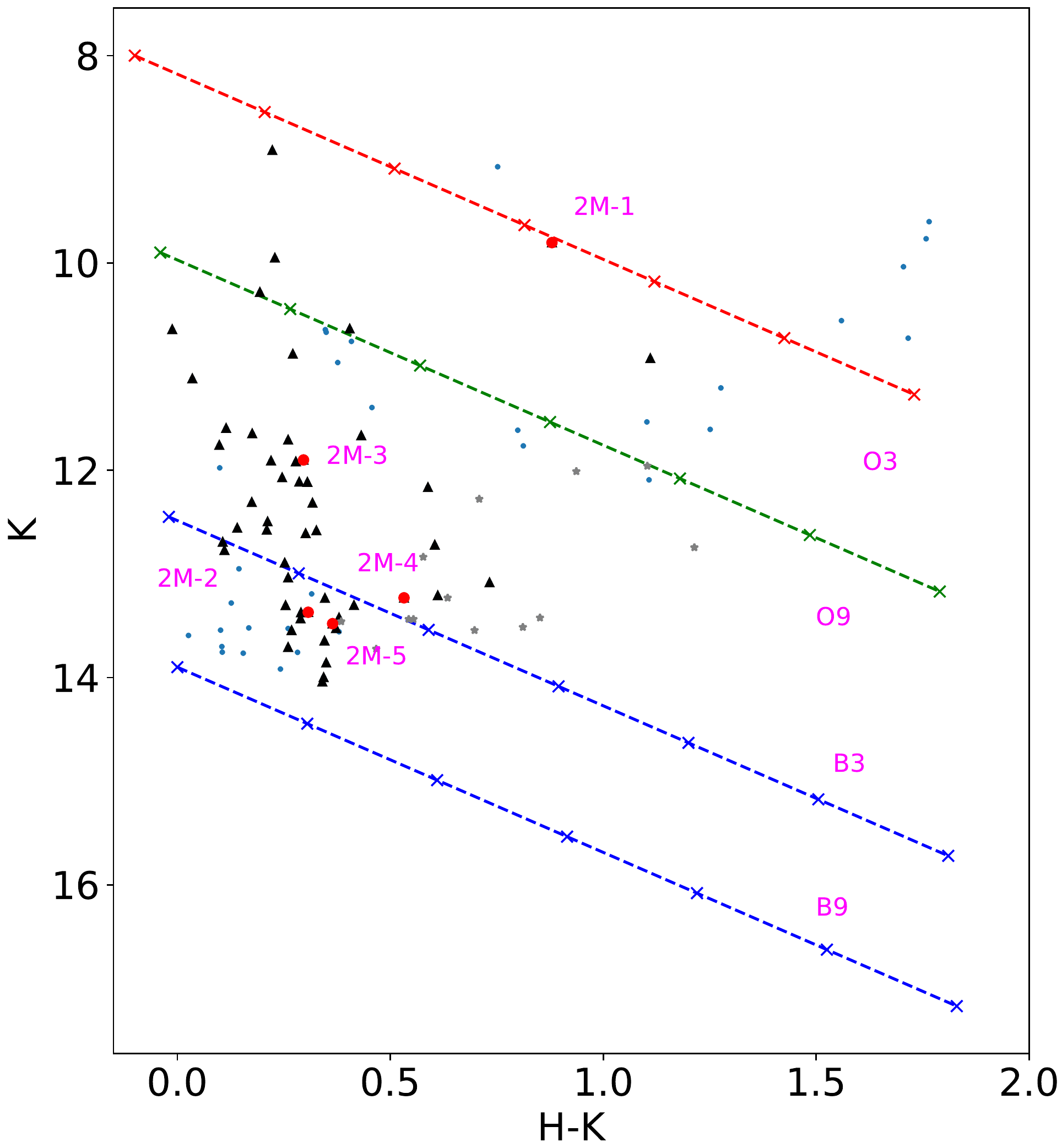}\includegraphics[width=\columnwidth]{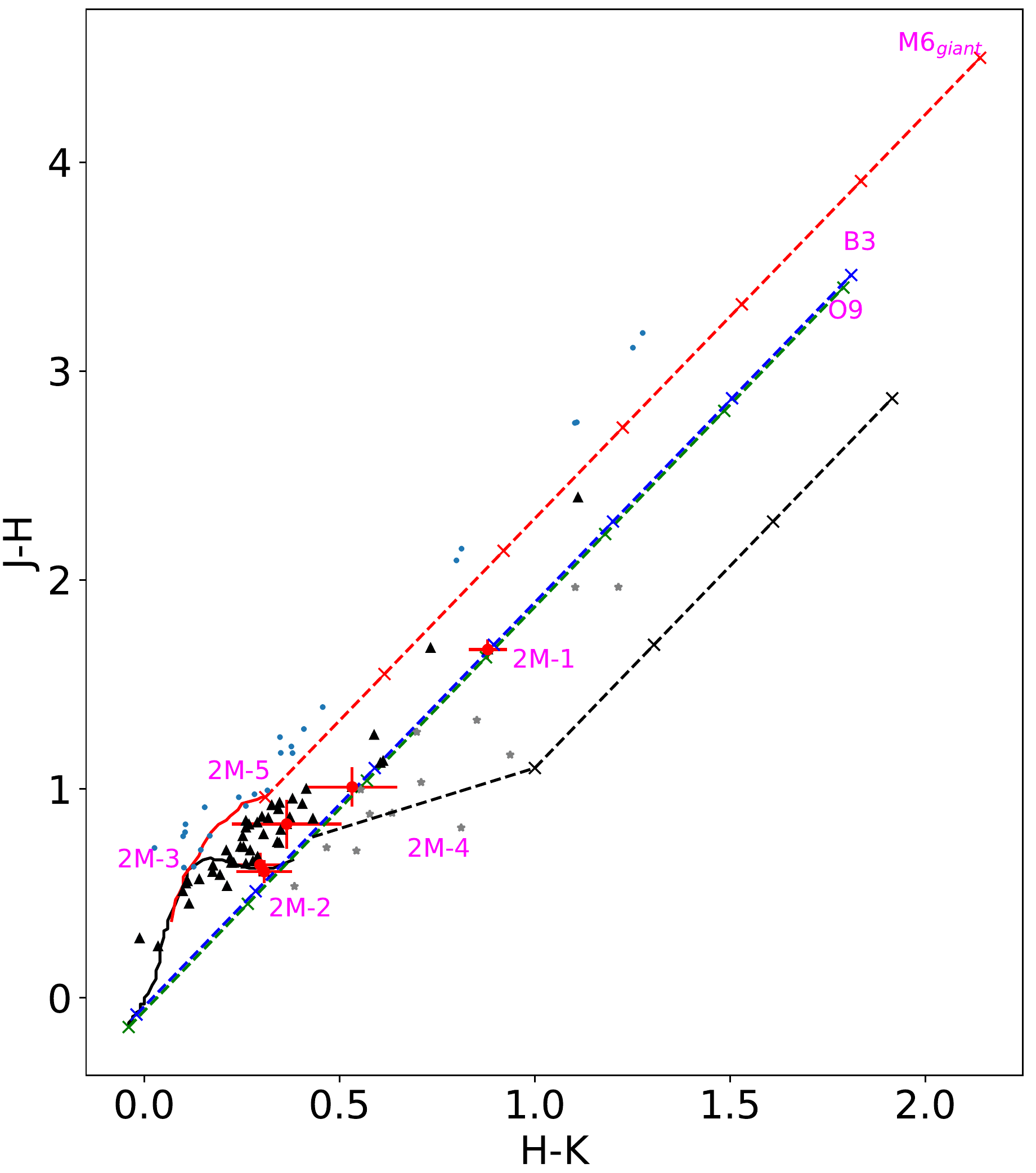}
   \caption{(left) The panel shows the K vs. (H--K)  color-magnitude diagram of the 2MASS near-infrared sources within a circle of radius $2\arcmin$ centered at G18.15. The reddening vectors of the massive stars (O3--B9) are plotted on the color-magnitude diagram. (right) The (J--H) vs. (H--K) color-color diagram of the 2MASS point sources being of spectral type earlier than B9 in the color-magnitude diagram is shown here. The solid black and red lines represent the loci of Class V (main sequence stars) and Class III (giant stars) objects respectively \citep{1988PASP..100.1134B,2004astro.ph..6003A}. The black long-dashed line represents the locus of the T-Tauri stars \citep{1997AJ....114..288M}. Reddening vectors corresponding to the different classes are also shown using dotted lines. The crosses on the reddening vectors are placed at an increasing interval of $A_{\text{V}}=5$. Sources (main sequence and giants) that are located within the reddening vectors of M6$_{\text{giant}}$ and O9 spectral classes are represented by black triangles, whereas pre-main sequence sources are denoted by gray stars. Any other objects are denoted by the cyan dots. The candidate ionizing stars are denoted by the red dots with respective error bars.}
          \label{fig:figure16}
    \end{figure*}

\begin{figure*}
	\includegraphics[width=\columnwidth]{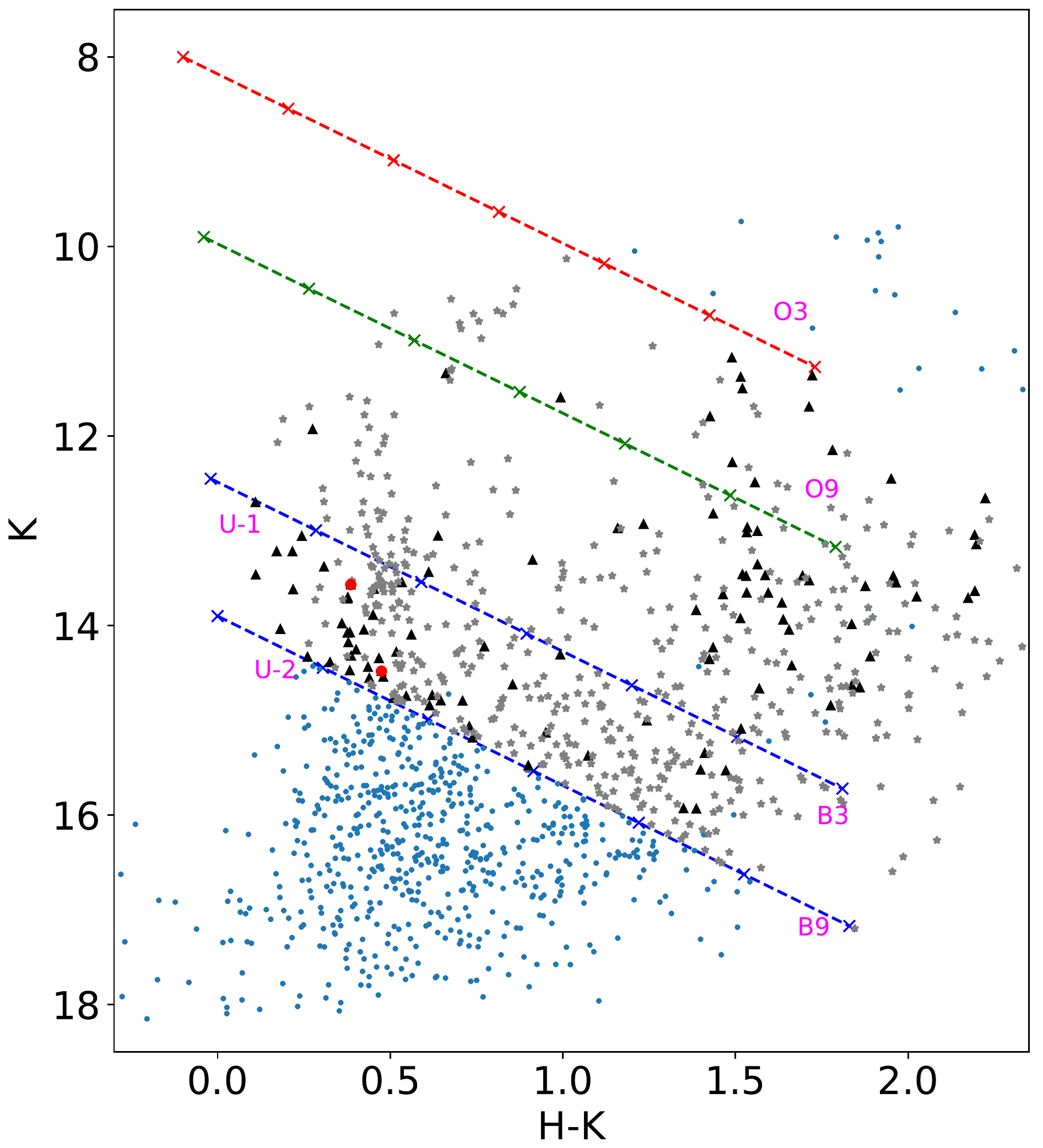}\includegraphics[width=\columnwidth]{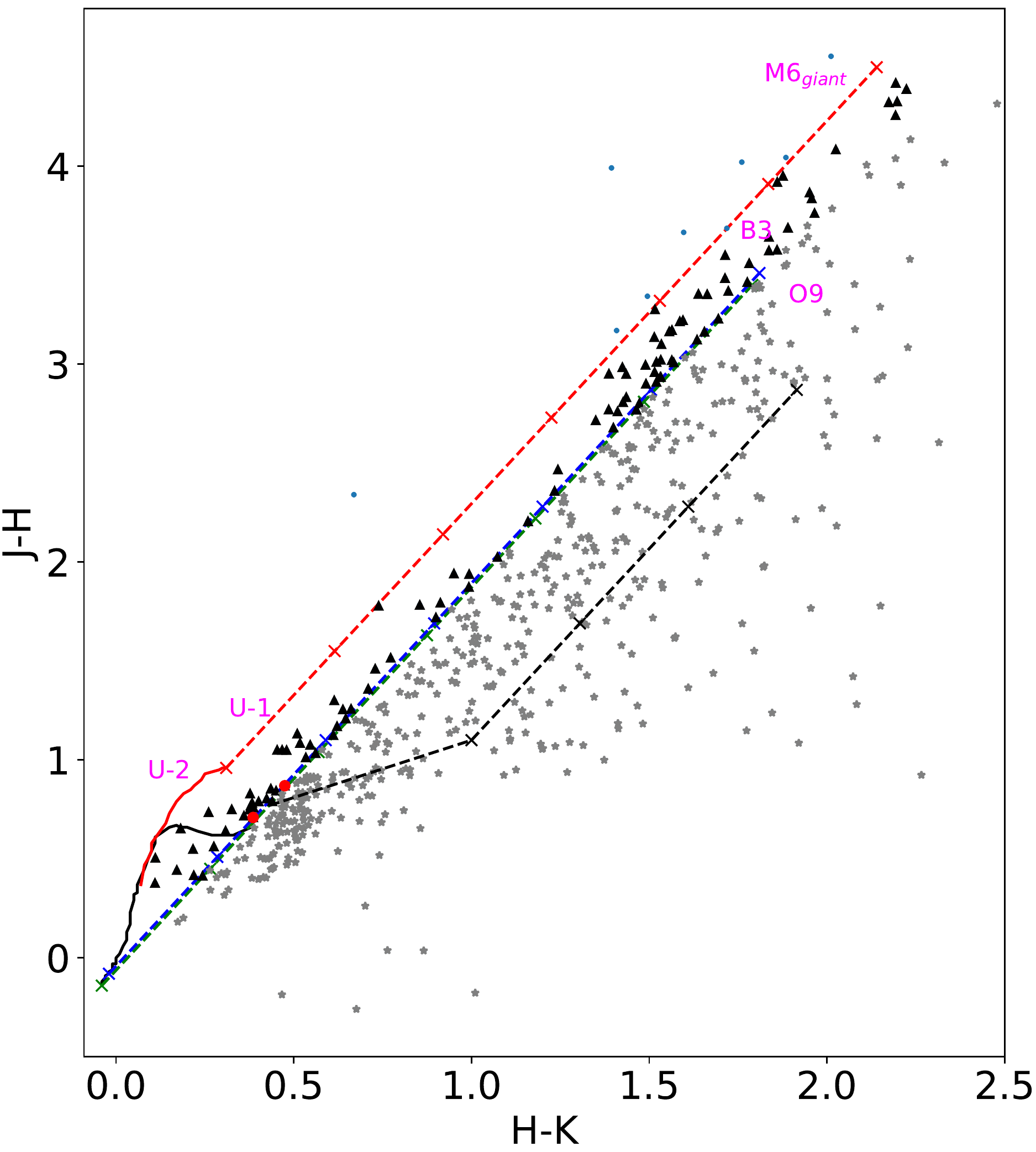}
    \caption{Same as Figure~\ref{fig:figure16} but for sources detected in the \textit{UKIDSS Galactic Plane Survey}.}
    \label{fig:figure16a}
\end{figure*}


\begin{figure}
	\includegraphics[width=\columnwidth]{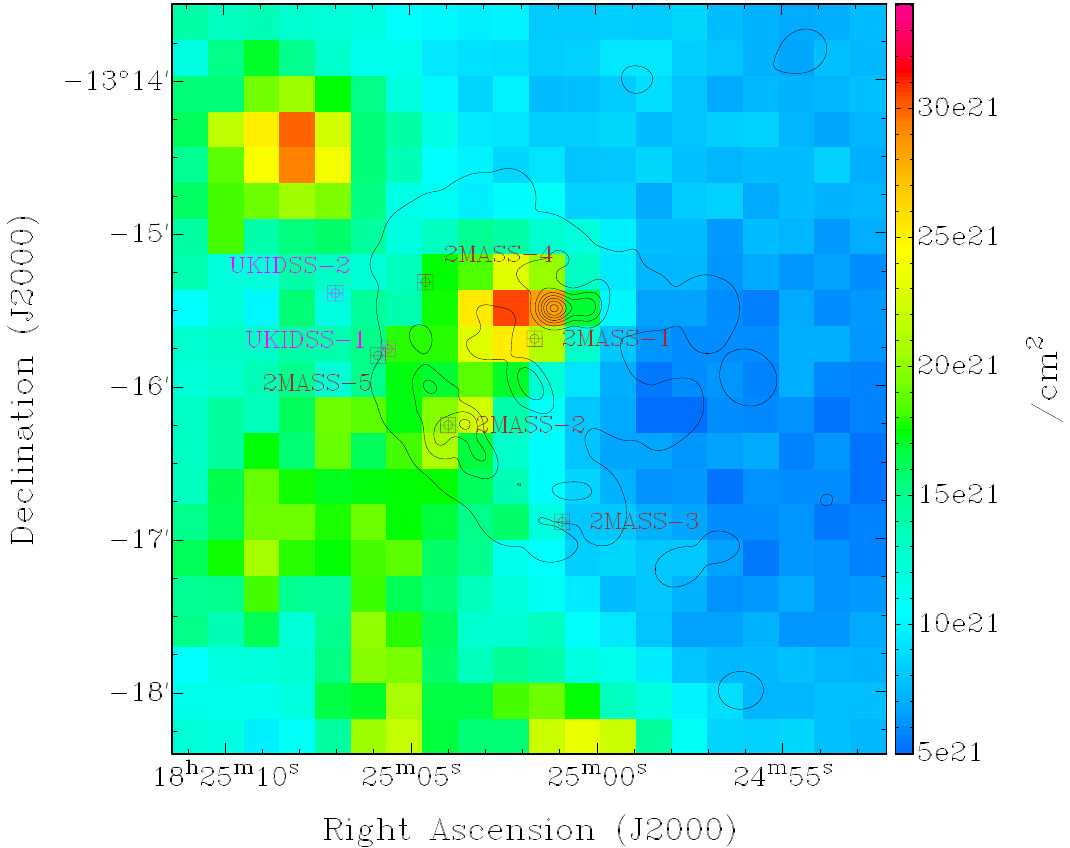}
    \caption{Locations of the candidate ionizing stars on the column density map estimated from the studies of cold dust.}
    \label{fig:figure17a}
\end{figure}

   \begin{figure*}
   \includegraphics[width=\columnwidth]{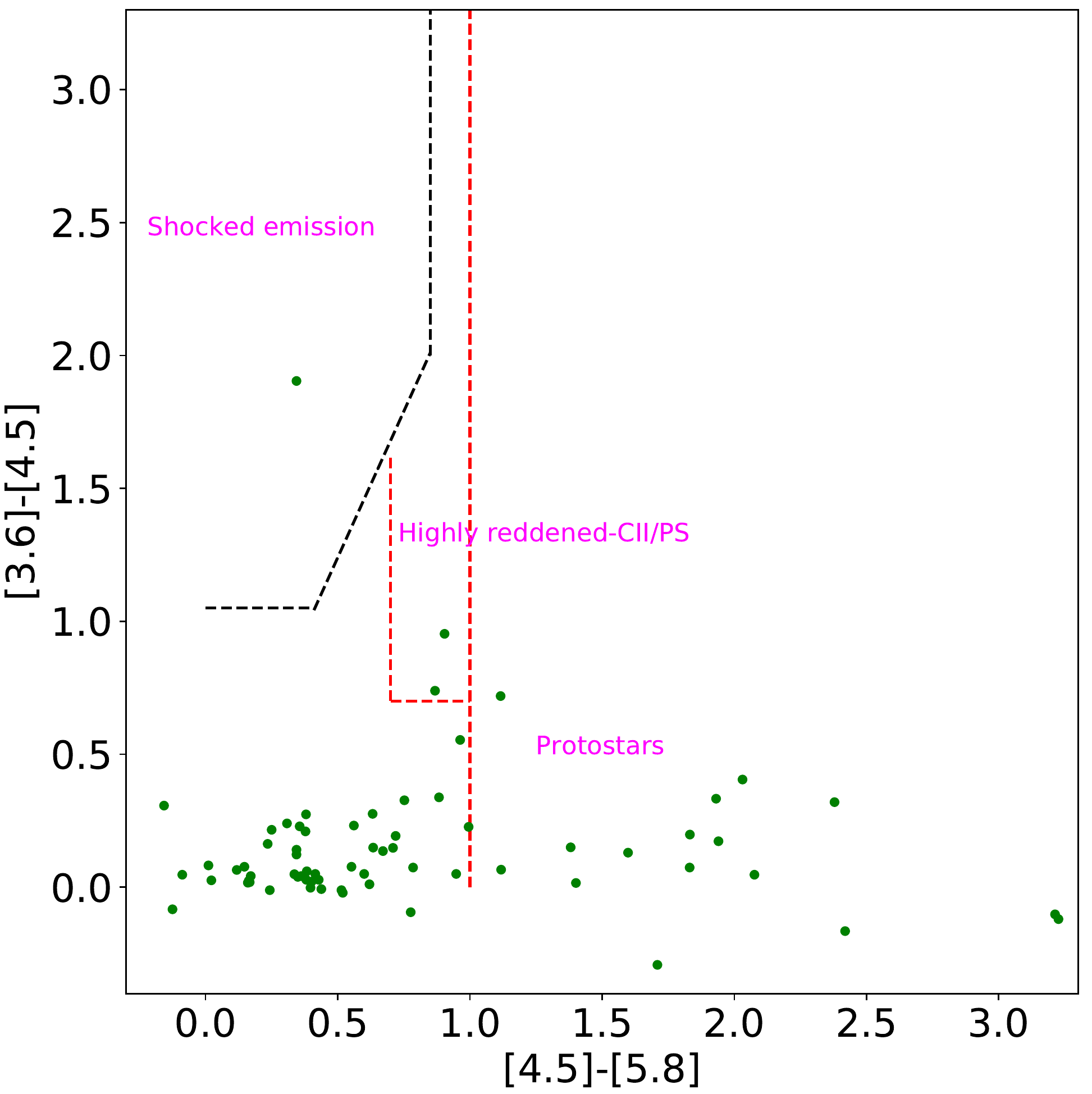}\includegraphics[width=\columnwidth]{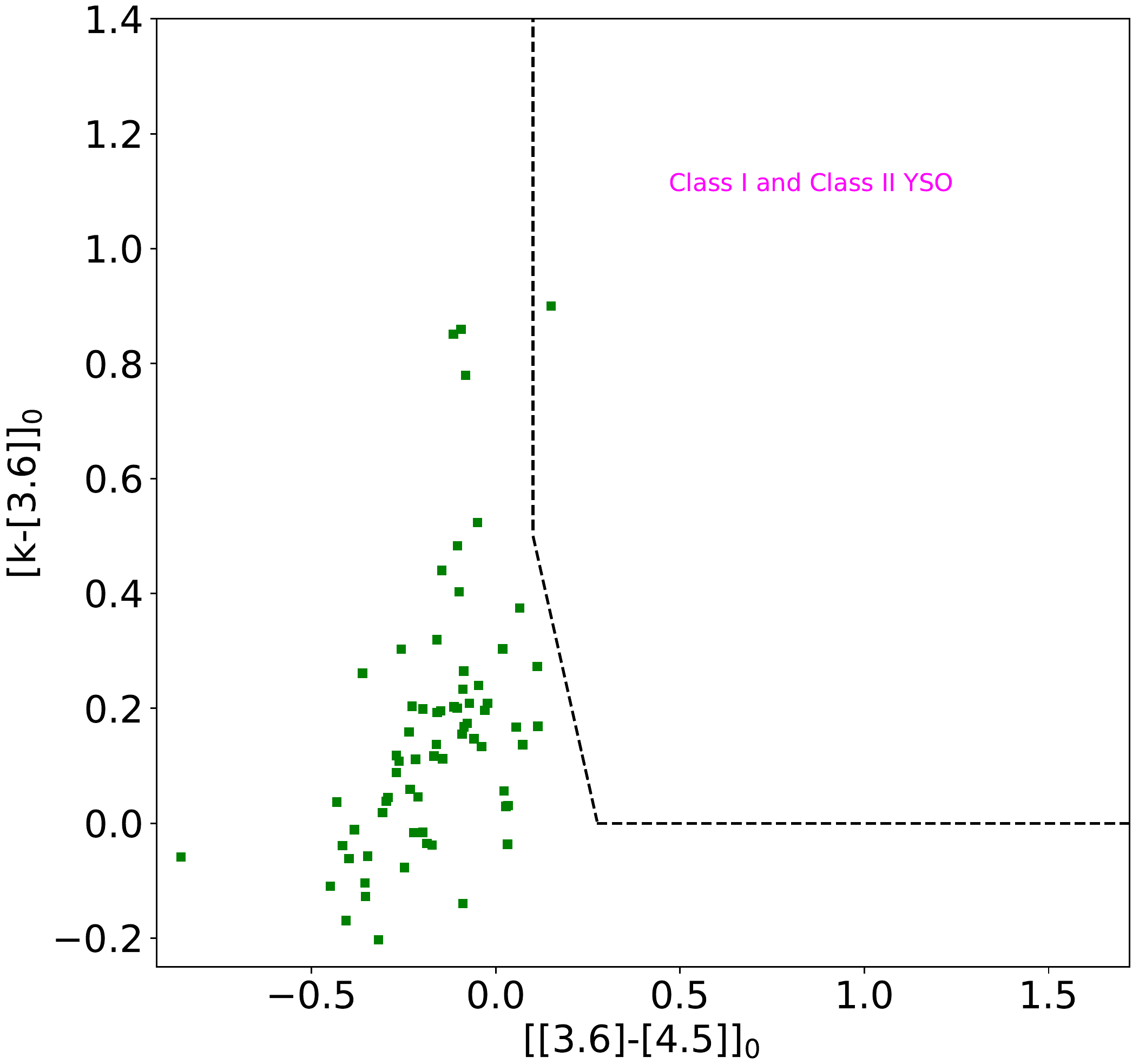}
   \caption{(left) The color-color diagram of the \textit{Spitzer}-GLIMPSE point sources within the searching radius centered at G18.15 is shown here. Sources that are detected in the IRAC-bands at 3.6, 4.5 and 5.8~$\mathrm{\mu m}$ simultaneously are plotted in the diagram. Regions belonging to different classes of objects are shown following the color conditions given in \citet{2008ApJ...674..336G}. (right) The panel shows the color-color diagram of the \textit{Spitzer}-GLIMPSE point sources detected in the J, H, and K bands of 2MASS along with 3.6 and 4.5-$\mathrm{\mu m}$ bands of IRAC simultaneously.}
          \label{fig:figure18}
    \end{figure*} 

\subsection{Orientation of the magnetic field lines}

We have utilized the 353-GHz \textit{Planck} dust polarization data to determine the mean orientation of the magnetic field lines in the vicinity of G18.15. Following the IAU convention (i.e., the position angle, $\theta_{\text{GAL}}=0\degr$ points
Galactic North but increases towards Galactic East), the $\theta_{\text{GAL}}$
values are derived using the relation,

\begin{equation}
    \theta_{\text{GAL}} = 0.5 \times \text{tan}^{-1}(-U/Q)
\end{equation}

Adopting the transformation relation of position angles \citep{1998MNRAS.297..617C}, the magnetic field orientation in Equatorial coordinates can be calculated as,

\begin{equation}
    \theta_{\text{B}} = \theta_{\text{GAL}} + \frac{\pi}{2} - \psi
\end{equation}

where $\psi$ is the transformation relation of the position angles in the Equatorial and Galactic systems at the position of each pixel. It is expressed as,

\begin{equation}
    \psi = \text{tan}^{-1}\left[\frac{\text{cos}(l-32.9\degr)}{\text{cos}(b)\text{cot}(62.9\degr)-\text{sin}(b)\text{sin}(l-32.9\degr)}\right]
\end{equation}

The mean orientation of the magnetic field is estimated to be at a position angle of $30.8\degr\pm2.6\degr$ as shown in Figure~\ref{fig:figure20}. It can be seen that the mean magnetic field is nearly perpendicular to the underlying central filament within G18.15. 

    
\subsection{Identification of the candidate ionizing stars}
\label{sec:subsection6}

In order to identify the candidate ionizing stars and YSOs towards G18.15, we have performed a photometric study of the near-infrared and mid-infrared point sources using data from the \textit{Two Micron All Sky Survey} (2MASS)\footnote{\url{http://vizier.u-strasbg.fr/viz-bin/VizieR-3?-source=II/246/out}}, \textit{UKIRT Infrared Galactic Plane Survey} (UKIDSS GPS)\footnote{\url{http://wsa.roe.ac.uk:8080/wsa/region_form.jsp}}, and \textit{Galactic Legacy Infrared Mid-Plane Survey Extraordinaire} (GLIMPSE)\footnote{\url{https://irsa.ipac.caltech.edu/data/SPITZER/GLIMPSE/}} surveys. We have searched for the candidate ionizing stars in the \textit{2MASS All-Sky Catalog of Point Sources} and \textit{UKIDSS GPS sixth archival data release} (UKIDSSDR6plus) catalogs using a circle of $2\arcmin$ radius centered at $\alpha=18^{\text{h}}25^{\text{m}}01^{\text{s}}\,,\delta=-13\degr16\arcmin02\arcsec$.

Next, two separate color-magnitude diagrams (CMD; K against H$-$K) are generated using the Bessell \& Brett homogenised system \citep{2001AJ....121.2851C} equivalent colors of the point sources assuming that all the sources are at the distance to G18.15. Although this process will lead to spurious sources that are misclassified as the OB-type stars on account of incorrect distances, these can be removed by constructing the color-color diagrams, and only selecting sources whose colors are consistent with that expected from the OB-type stars. Thus, we selected the OB-type candidates from the CMD and plotted them separately in respective color-color diagrams (CCD; H$-$K against J$-$H). We have also assumed the interstellar reddening law of \citet{1985ApJ...288..618R} ($A_{\text{J}}$/$A_{\text{V}}$ = 0.282; $A_{\text{H}}$/$A_{\text{V}}$ = 0.175 and $A_{\text{K}}$/$A_{\text{V}}$ = 0.112) to draw the reddening vectors, where the crosses are placed at an increasing interval of $A_{\text{V}}$ = 5. The CMD and CCD from 2MASS are shown in the left and right panels of Figure~\ref{fig:figure16}, and Figure~\ref{fig:figure16a} shows the same for UKIDSS. 

Following our analysis, we have identified 2 and 5 candidate ionizing stars from the UKIDSS and 2MASS surveys respectively. Table~\ref{tab:table4} lists all 7 candidates identified using the near-infrared photometry, and they are shown in Figure~\ref{fig:figure17a}. As an added consistency check, we determined the hydrogen column density ($N(\text{H}_2)$) expected at the coordinates of the candidate stars from the visual extinction ($A_{\text{V}}$) measured in the color-color diagram using \citet{1978ApJ...224..132B},

\begin{equation}
    \label{eq:equ19a}
    {N}(\text{H}_2) = \frac{{A}_{\text{V}}}{5.3 \times 10^{-22}}\, \text{cm}^{-2}
\end{equation}

The respective $N$(H$_2$) values from eq.~\ref{eq:equ19a} and locations of the candidate stars in Figure~\ref{fig:figure17a} are also listed in Table~\ref{tab:table4}. We find that there is broad consistency between the column density measured from the far-infrared map and that estimated by the visual extinction, with discrepancies attributed to the uncertainty in the dust opacity and other variations in the $N$(H$_2$) to $A_{\text{V}}$ ratio \citep{1995A&A...293..889P,2009MNRAS.400.2050G}.

This gives confidence that the sources listed in Table~\ref{tab:table4} are indeed OB-type stars that are associated with the G18.15 region. However, it is to be noted that only two stars are of spectral type earlier than B0. Figure~\ref{fig:figure17a} shows that 2 (2MASS-1 and 2) out of the 7 candidate ionizing sources are located near the continuum peaks of G18.15, whereas 4 out of the remaining 5 sources are located at the periphery of the central dust filament with the last source (UKIDSS-2) at a location slightly away from the region showing radio continuum emission. Thus, while 2MASS-2 is the likely source for the southern continuum peak, the ionizing stars for the northern peak are not detected. However, our analysis does not preclude the existence of a population of ionizing stars in the IRDC that are not detected at near-infrared wavelengths on account of significant extinction. This is consistent with the overall ionizing photon flux in the HII region, which is significantly higher than that produced by a single O9 star.

\subsection{YSOs associated with the G18.15}
\label{sec:subsection7}

   \begin{figure*}
   \includegraphics[width=\columnwidth]{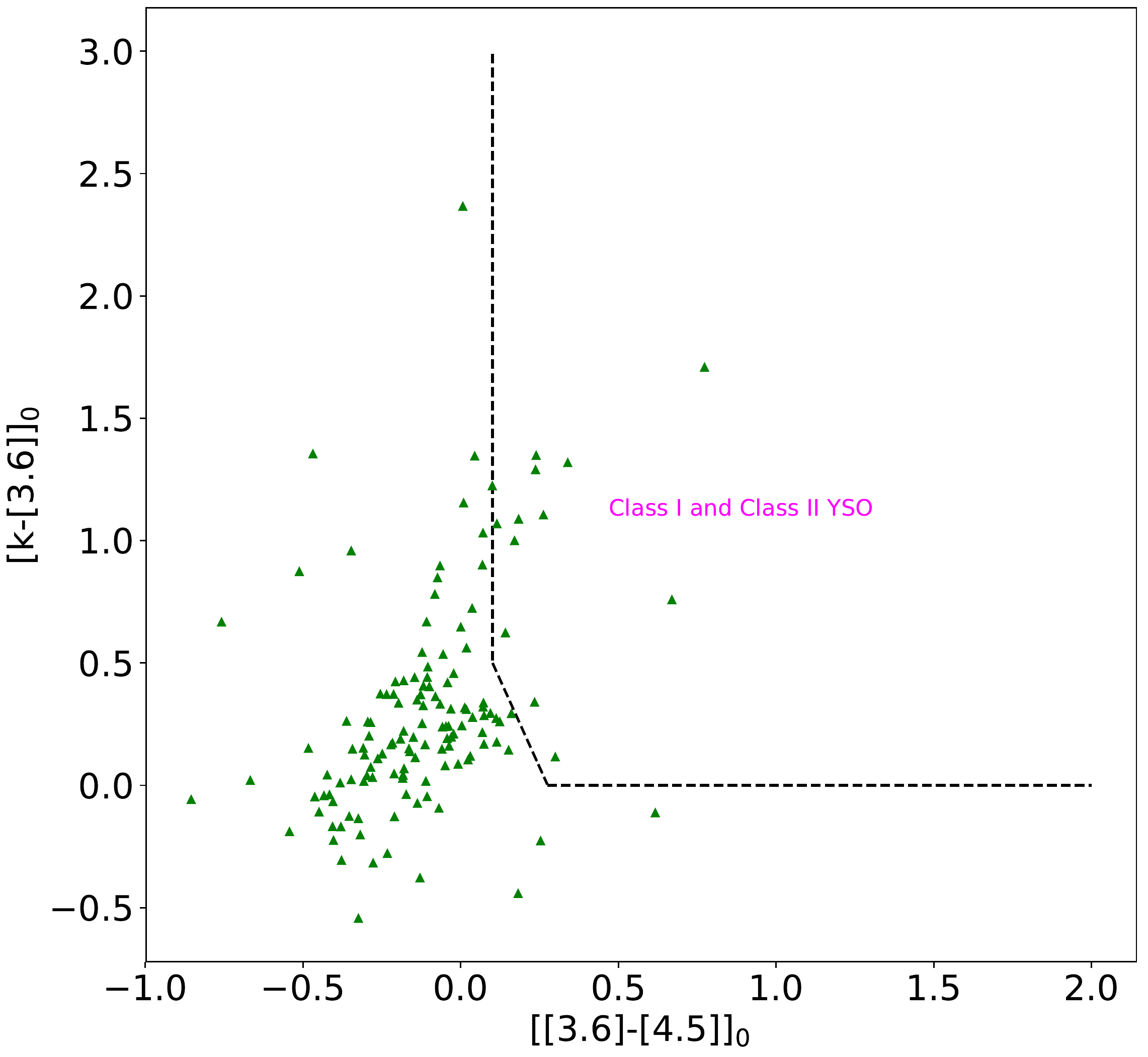}
   \includegraphics[width=\columnwidth]{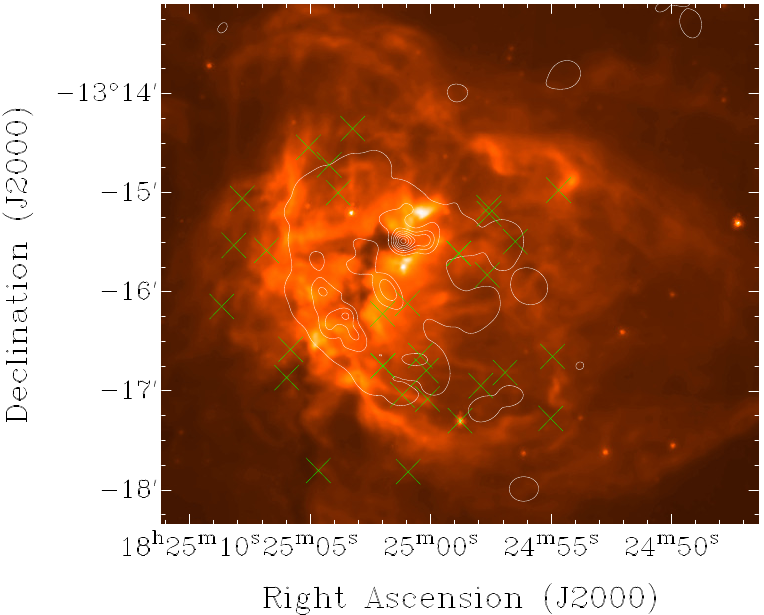}
   \caption{(left) The color-color diagram of the \textit{Spitzer}-GLIMPSE point sources detected in the J, H, and K bands of UKIDSS along with 3.6 and 4.5-$\mathrm{\mu m}$ bands of IRAC simultaneously is shown here. The JHK-colors of UKIDSS are converted to the JHK-colors of 2MASS following \citet{2001AJ....121.2851C} before constructing the color-color diagram. (right) The panel shows the locations of the YSOs (green crosses) that are detected from the GLIMPSE, 2MASS and UKIDSS surveys on the 8-$\mu$m warm dust emission map towards G18.15.}
          \label{fig:figure19}
    \end{figure*} 

In order to identify YSO candidates, we have searched the \textit{Spitzer}-GLIMPSE online database within the same circular region ($2\arcmin$ radius centered at $\alpha=18^{\text{h}}25^{\text{m}}01^{\text{s}}\,,\delta=-13\degr16\arcmin02\arcsec$) used for searching the candidate ionizing stars and found a total of 369 mid-infrared sources inside our region of interest. Due to the nebulosity affecting the IRAC 8.0-$\mu$m band, most of the sources are not detected in all four bands of the IRAC camera. Thus, we have adopted IRAC three-band and 2MASS-IRAC five-band classification schemes as described in \citet{2008ApJ...674..336G} to classify the YSOs. These classification schemes are based on the [4.5]--[5.8] color, as it is less contaminated by the dust extinction than the 3.6-$\mu$m emission-based colors \citep{ 2005ApJ...629..881H}.



Following the IRAC three-band classification scheme, we have detected 18 protostar candidates within G18.15, of which 2 are likely to be highly reddened Class II objects. We have also detected an additional YSO using the 2MASS-GLIMPSE five-band classification scheme. The corresponding CCDs are shown in the left and right panel of Figure~\ref{fig:figure18} respectively. 

Since UKIDSS is a much deeper survey compared to 2MASS, we have also searched for YSOs using the NIR and MIR colors of UKIDSS and GLIMPSE. We first identified counterparts of the GLIMPSE point sources in the UKIDSS GPS using a $0.5\arcsec$ matching radius. The UKIDSS colors were then converted to equivalent 2MASS colors following which the five-band classification scheme was used to identify YSOs. A total of 13 Class I or II objects are detected following this method, out of which 11 were undetected in the IRAC three-band and 2MASS-IRAC five-band classification schemes. Overall, we have detected 30 YSO using all aforementioned classification schemes. 

The left and right panels of Figure~\ref{fig:figure19} show the de-reddened CCD using the GLIMPSE and UKIDSS surveys and locations of detected YSOs towards G18.15. We find most of the YSOs to be located towards the edges of the HII region, including the IRDC extending beyond the radio emission, with only a couple of YSOs towards the central filamentary structure. As with the case of ionizing stars, a significant fraction of YSOs in the central IRDC is likely to be undetected because of high extinction.

\begin{figure*}
	\includegraphics[width=\columnwidth]{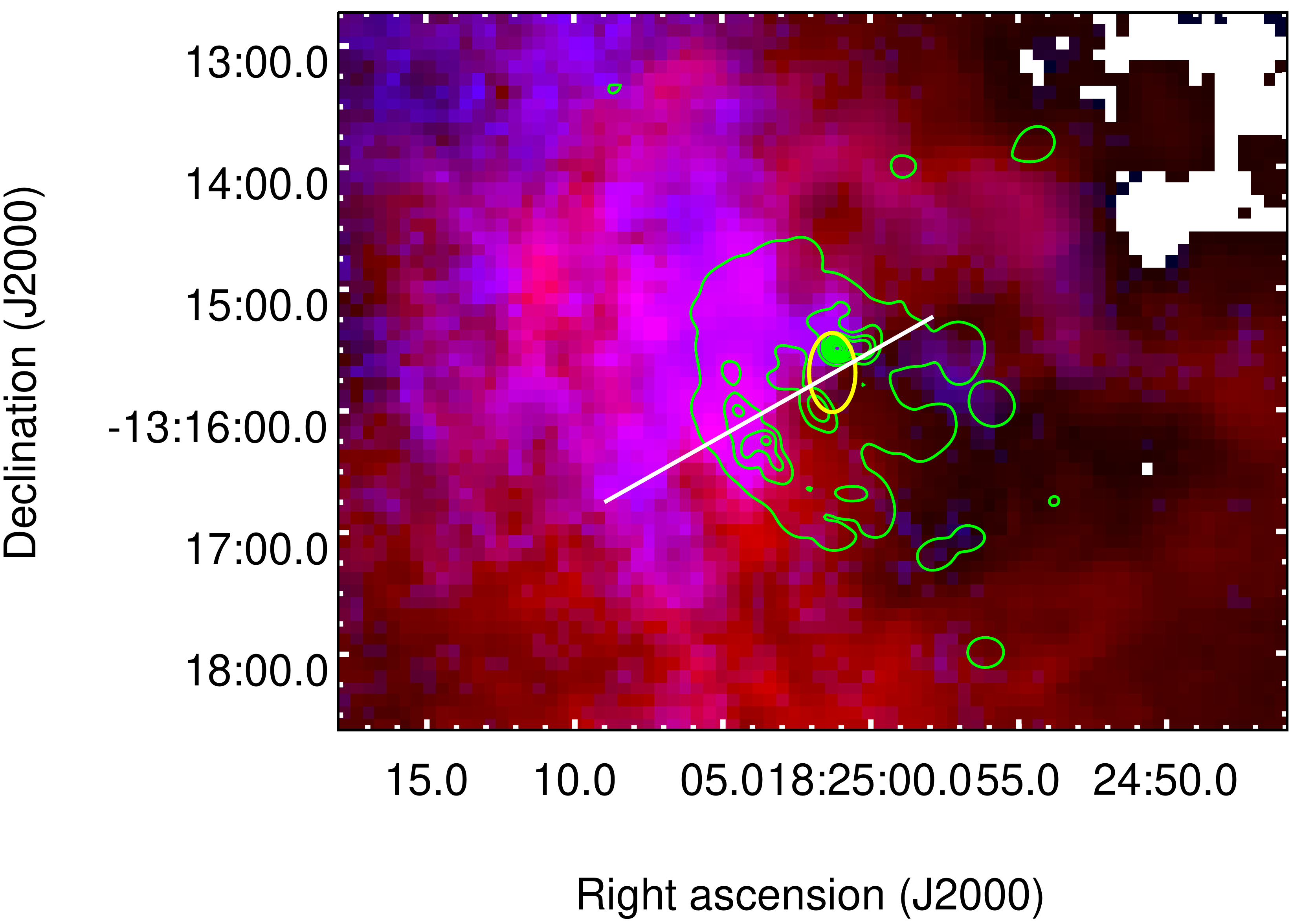}\includegraphics[width=\columnwidth]{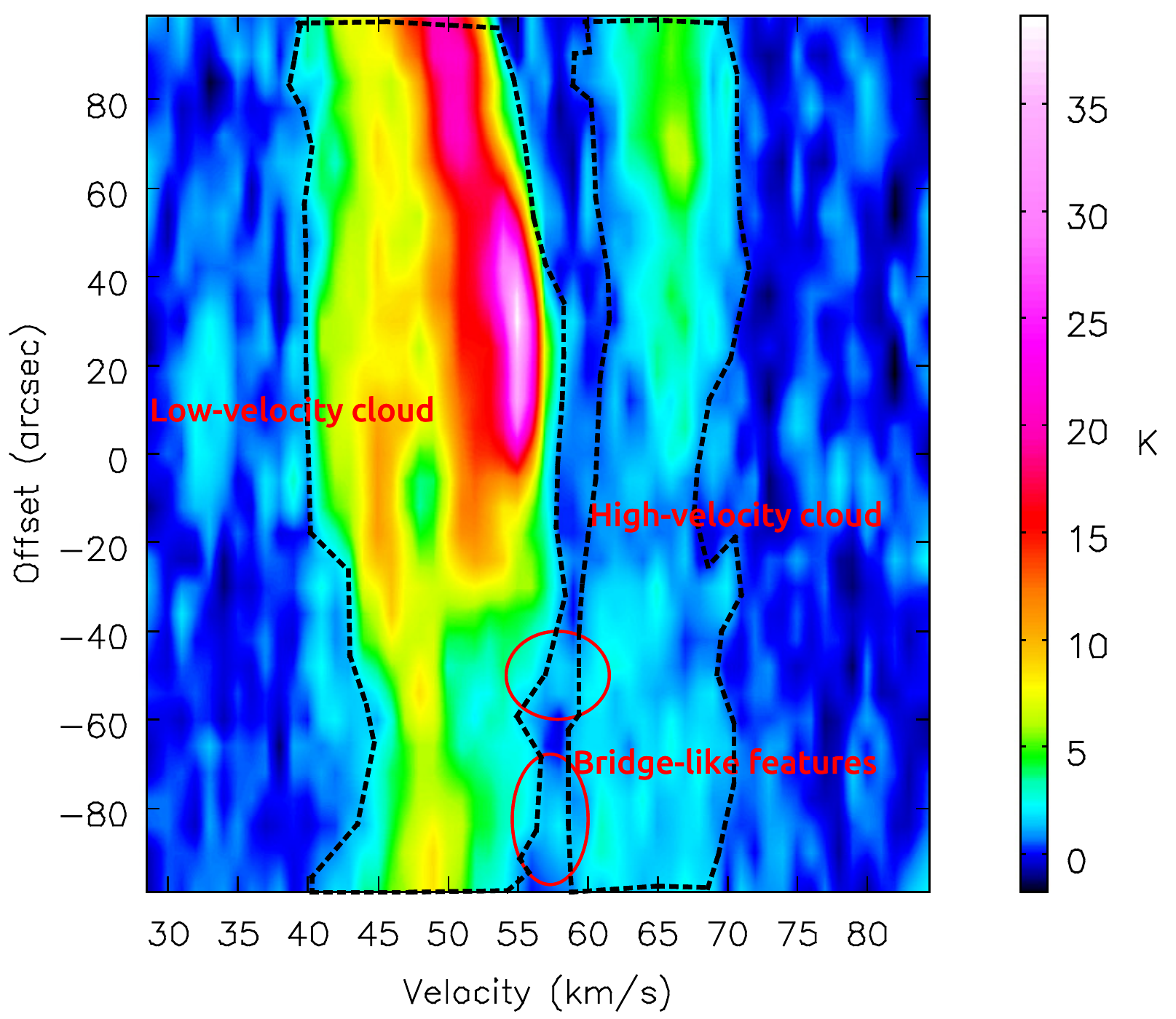}
    \caption{(left) The $^{12}$CO ($J$=3$-$2) intensity integrated within the velocity 39 to 55 km s$^{-1}$ (red) and 56 to 74 km s$^{-1}$ (blue) are shown in a two color composite image. The white solid line shows the cut along which the position-velocity diagram is extracted. The radio continuum contours in green are identical to those in Figure~\ref{fig:moment0}. (right) The panel shows the position-velocity diagram extracted along the cut shown in the left panel. The ``broad-bridge'' like features are visible, and are encircled in red. An approximate position of the feature located at an offset $\approx-50\arcsec$ is shown in a yellow ellipse in the left panel.}
    \label{fig:pv_cut}
\end{figure*}

\begin{figure*}
\centering
	\includegraphics[width=15cm]{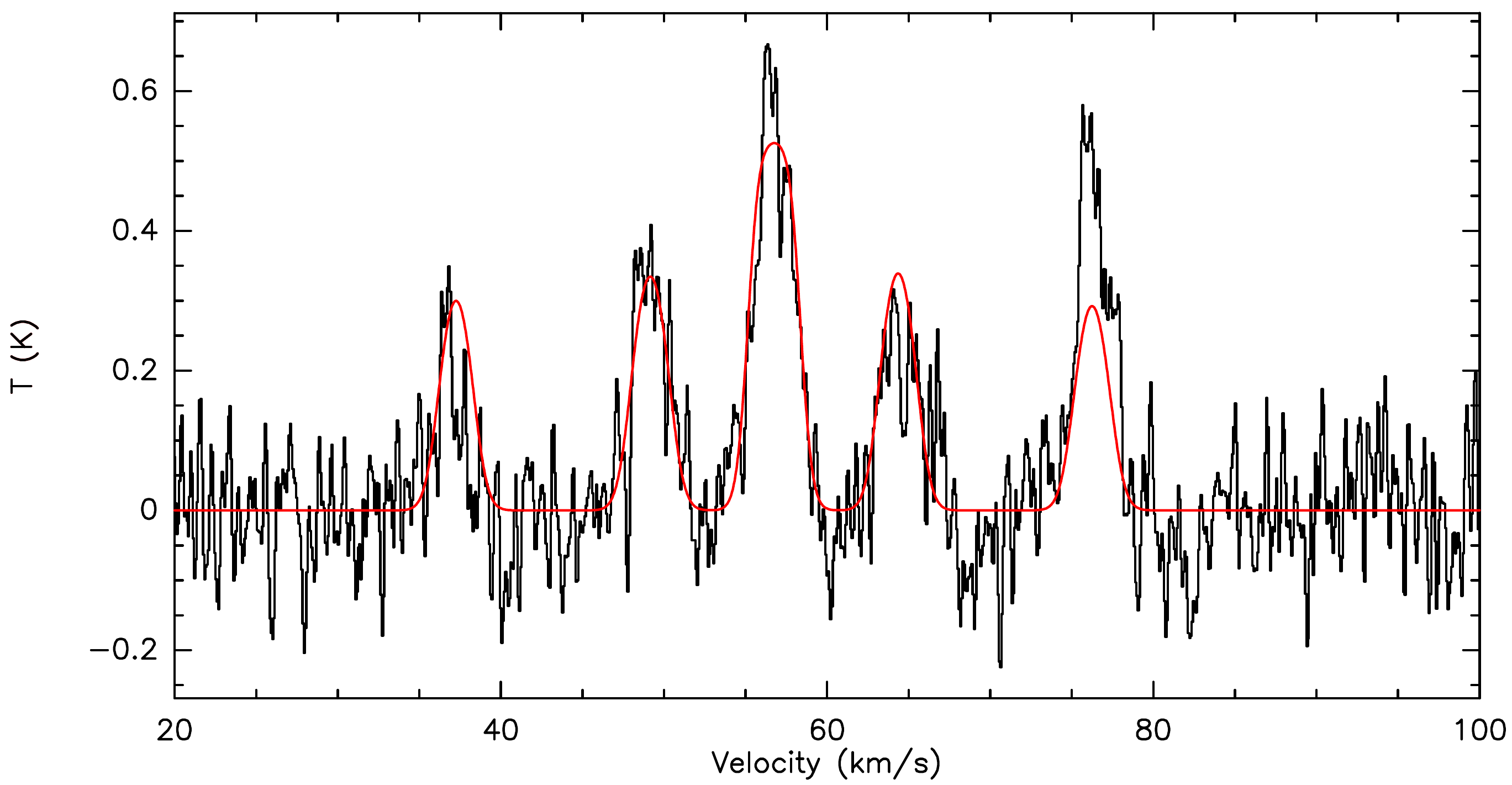}
    \caption{The figure shows the NH$_{3}$(1,1) inversion spectrum towards G18.15. The line peaks at 56.8 $\pm$ 0.03~km~s$^{-1}$ matching the velocity of the compressed layer.}
    \label{fig:nh3}
\end{figure*}

\begin{figure}
    \centering
    \includegraphics[width=\columnwidth]{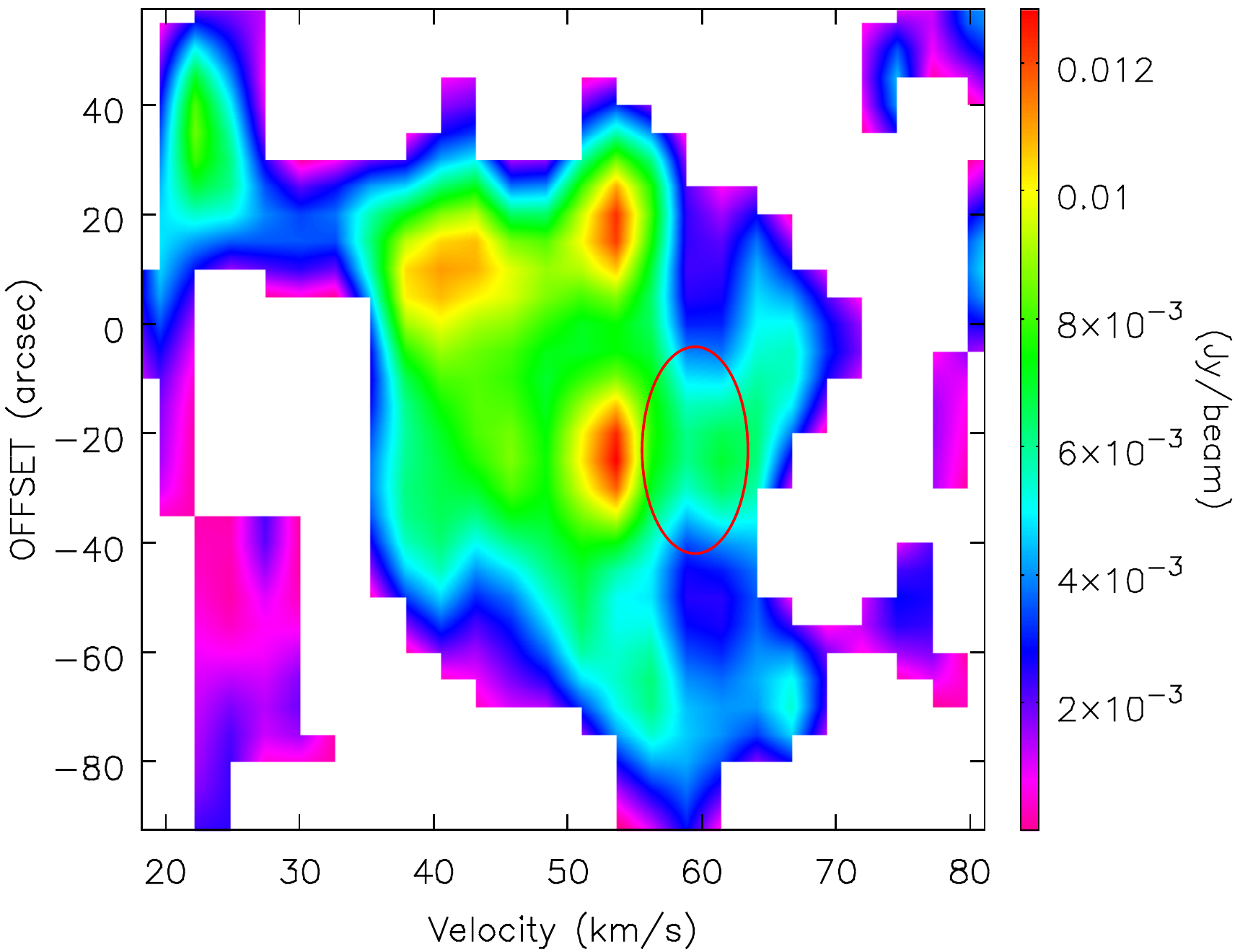}
    \caption{The figure shows the position-velocity diagram generated using the RRL emission towards G18.15. The diagram is extracted along the same locus as shown in the left panel of Figure~\ref{fig:pv_cut}. The highlighted ``broad-bridge'' feature is encircled in red.}
    \label{fig:pv_rrl}
\end{figure}

\section{Discussion}
\label{disc}
\subsection{Age of the region}

The age of G18.15 can be estimated using the observed properties of the HII region. Under the assumption that the HII region is expanding into a homogeneous medium, the Stromgren radius ($R_{\text{s}}$) is given by the following expression,

\begin{equation}
    \label{eq:equ20}
    {R}_{\text{s}} = \left(\frac{3\,{N}_{\text{Ly}}}{4\, \pi \, {n}_0^2\,\alpha_{\text{B}}}\right)^{1/3}
\end{equation}
where $\alpha_{\text{B}}$ is the radiative recombination coefficient assumed to be 2.6 $\times$ 10$^{-13}$~cm$^3$~s$^{-1}$ \citep{1989agna.book.....O}, and $n_0$ is the number density of atomic hydrogen, which can be derived from the column density map following $n_0$ $=$ $3N$(H$_2$)/$2R$. Here, $R$ is the radius of the ionized clump. We can estimate the dynamical age, $t_{\text{dyn}}$, of the HII region based on a simple model of an expanding HII region in a homogeneous medium \citep{1980pim..book.....D} as,

\begin{equation}
    \label{eq:equ21}
    {t}_{\text{dyn}} = \left[\frac{4}{7}\,\frac{{R}_{\text{s}}}{{c}_{\text{i}}}\right]\,\left[\left(\frac{{R}}{{R}_{\text{s}}}\right)^{7/4}-1\right]
    \label{eq:age}
\end{equation}
where $c_{\text{i}}$ is the sound speed in the ionized medium and is assumed to be 10~km~s$^{-1}$. The radius of the HII region ($R$) is estimated to be 1.31~pc considering the radio continuum emission above $3\sigma$ level, and the mean value of hydrogen column density calculated using the same area obtained from the radio continuum is equal to $1.43\times10^{22}$~cm$^{-2}$. Consequently, $n_0$ and $R_{\text{s}}$ are estimated to be $5.3\times10^3$~cm$^{-3}$ and 0.186~pc respectively. Since the size of the HII region is larger than that of the Stromgren sphere, the expansion of G18.15 is currently pressure-driven. Using eq.~\ref{eq:age}, we derive the dynamical age of G18.15 to be 0.31~Myr.

\subsection{Signatures of a CCC in G18.15}

The characteristic observational features of a CCC are the overlapping distribution of the colliding clouds, the presence of two velocity components in the CO spectra \citep{1992PASJ...44..203H, 2018PASJ...70S..58T}, and the ``broad-bridge'' feature connecting the velocity peaks at the intermediate velocity range \citep{2015MNRAS.450...10H}. G18.15 manifests many of these characteristics that suggest that it is a site of a CCC event.

The RRL data show two distinct velocities with a region of intermediate velocity separating them spatially. In addition, the spectrum of $^{12}$CO molecular emission also shows two velocity components connected by a narrow plateau-like emission profile at the intermediate velocities (Figure~\ref{fig:figure8}). The velocity peaks of the CO emission are seen to be in close correspondence with that of the RRLs, indicating that both the CO and RRL emission are tracing the same molecular clouds.

To further test the CCC hypothesis, we have searched for the presence of a ``broad-bridge'' like feature connecting the two molecular clouds in the position-velocity (PV) diagram. The right panel of Figure~\ref{fig:pv_cut} shows the PV diagram from the COHRS data generated across the white line shown in the left panel of the exact figure wherein the bridging features are visible. An approximate location of these features is shown using a yellow ellipse on the white line. However, the observed ``broad-bridge'' features in the CO ($J$=3$-$2) PV diagram are much weaker than what is observed in the simulations of \citet{2015MNRAS.450...10H}. This is surprising since the bridge feature must be strong at a phase where the cloud-cloud interaction has triggered star formation in the colliding clouds. Moreover, the $J$=3$-$2 transition of CO is expected to be a better tracer of dense, compressed gas compared to the (1$-$0) transition in the \citet{2015MNRAS.450...10H} simulations. However, the NH$_3$ spectrum towards the region from the \textit{Red MSX Survey} \citep{2011MNRAS.418.1689U} using the \textit{Robert C. Byrd Green Bank telescope} shows strong emission at the intermediate velocity of 56.8 $\pm$ 0.03 km s$^{-1}$ (Figure~\ref{fig:nh3}). Since NH$_3$ is an excellent tracer of the dense gas, this strongly suggests presence of a significant amount of dense gas in the collision interface. This agrees with the simulations of \citet{2021MNRAS.506..775P} where molecules tracing dense gas were observed to have strongly enhanced emission in the shock-compressed layer between colliding clouds. The smooth velocity gradient in the RRL emission also suggests that the bridge feature may be visible in a PV diagram constructed from the RRL data. This is indeed the case as seen in Figure~\ref{fig:pv_rrl}, wherein the bridge feature is as prominent as the component at high velocity. The prominent bridge feature in the RRL data along with the presence of dense molecular gas at the intermediate velocity add to the evidence of G18.15 being a possible site of a cloud-cloud collision.

Moreover, the far-infrared emission and mid-infrared absorption reveal a dense filament that passes through the central regions of G18.15 extending further to the south, and the mean magnetic field orientation is estimated to be perpendicular to the central filament within the HII region. Although the magnetic field geometry by itself cannot be taken to be evidence for a CCC, it adds to the results obtained from the PV diagram, and suggests a CCC hypothesis behind the formation of G18.15.

In addition to the massive stars inhabiting G18.15, the detection of several YSOs suggests that star formation is actively ongoing in the cloud complex. As mentioned earlier, the presence of extended emission at mid-infrared wavelengths and the high extinction as seen by the dark features at 8.0~$\mu$m suggest a significant undetected population of YSOs and massive stars.

\section{Summary and Conclusions}
\label{sec:section5}

In this paper, we have performed a multi-wavelength analysis of G18.15. Our major findings are listed below:
\begin{itemize}
    \item Using high resolution uGMRT data, we have detected multiple radio continuum peaks within G18.15. The H169$\alpha$ and H170$\alpha$ RRLs are also detected towards G18.15, and are stacked to increase the SNR of the final map. The RRL data reveal two distinct velocity components in G18.15.
    \item The electron temperature is determined using the RRL and radio continuum data and is found to be 5200$-$9500~K at the location of the continuum peaks, which is consistent with the average electron temperature determined by previous studies using single-dish telescopes.
    \item The COHRS data reveal two molecular clouds with LSR velocities of 53.4 and 66.7~km~s$^{-1}$ respectively. A narrow plateau-like emission connects these two clouds at the intermediate velocities. The NH$_{3}$(1,1) spectrum shows a strong peak at $56.8$~km~s$^{-1}$ located within the intermediate velocity range.
    \item The position-velocity diagrams of $^{12}$CO and RRL reveal the presence of ``broad-bridge'' features. 
    \item The Herschel far-infrared emission and GLIMPSE 8.0~$\mu$m absorption reveal an IRDC filament with G18.15 at the northern edge.
    \item The mean orientation of the magnetic field lines is found to be perpendicular to the HII region.
    \item Using 2MASS and UKIDSS photometric data, star formation activity is investigated, and two O9 stars are identified towards G18.15. Although 30 YSOs have been detected towards this region, the extended emission at mid-infrared wavelengths and high extinction suggest the presence of an undetected population of massive stars and YSOs.
    \item The dynamical age of G18.15 is estimated to be 0.3~Myr approximately.
\end{itemize}

Following the observational evidences presented in this paper, we conclude that G18.15 is very likely to be formed due to a cloud-cloud collision event roughly 0.3~Myr ago.

\section*{Acknowledgements}

The authors thank the referee for the detailed and constructive comments, which helped to improve the clarity of the paper. We also thank the Max Planck Society for funding this research through the Max Planck partner group initiative. D.~V.~L. acknowledges the support of the Department of Atomic Energy, Government of India, under project no. 12-R\&D-TFR-5.02-0700. J.~D. is thankful to Dr.~C.~H.~Ishwara~Chandra and Dr.~Ruta~Kale for their valuable inputs regarding the data reduction. The authors also thank Mr.~Jean~Baptiste~Jolly from the Chalmers University of Technology and the Nordic ALMA Regional Center node for providing the Line-Stacker package. This research has made use of the SIMBAD database, operated at CDS, Strasbourg, France. This research has also utilized NASA’s Astrophysics Data System and CDS' VizieR catalog access tool.


\appendix

\section{Calculation of the electron temperature}
\label{sec:appendix}
We derive the electron temperature under the local thermodynamic equilibrium (LTE) conditions. Under LTE and Rayleigh-Jeans limit, the observed specific intensity takes the following form,

\begin{equation}
   {I}_{\text{cont}}(\nu)=\frac{2\,{k}_{\text{B}}\,{T}_{\text{e}}\,\nu^2}{{c}^2}\: \left( 1-{e}^{\tau_{\text{cont}}(\nu)} \right)
   \label{eq:eq2}
\end{equation}
where $\nu$ is the frequency, $k_{\text{B}}$ is the Boltzmann constant, $T_{\text{e}}$ is the electron temperature, and $\tau_{\text{cont}}(\nu)$ is the optical depth of the medium for continuum emission. Rearranging the terms, eq.~\ref{eq:eq2} can also be written as,

\begin{equation}
  \label{eq:eq3}
  \tau_{\text{cont}}(\nu) = - \ln \left( 1 - \frac{{c}^2\,{I}_{\text{cont}}(\nu)}{2\,{k}_{\text{B}}\,{T}_{\text{e}}\,\nu^2} \right)
\end{equation}
Now, from the definition of brightness temperature, the continuum temperature ($T_{\text{cont}}$) can be estimated as,

\begin{equation}
    \label{eq:eq4}
    {T}_{\text{cont}} = \frac{{c}^2\, {I}_{\text{cont}}(\nu)}{2\, {k}_{\text{B}}\, \nu^2}
\end{equation}
Similarly, the brightness temperature of the spectral line ($T_{\text{line}}$) is,

\begin{equation}
    \label{eq:eq5}
    {T}_{\text{line}} = \frac{{c}^2\, {I}_{\text{line}}(\nu)}{2\, {k}_{\text{B}}\, \nu^2}
\end{equation}
for any given pixel in the image. Thus, the line-to-continuum intensity ratio in a single pixel is,

\begin{equation}
  \label{eq:eq6}
  \frac{{I}_{\text{line}}(\nu)}{{I}_{\text{cont}}(\nu)} = \frac{{I}(\nu)-{I}_{\text{cont}}(\nu)}{{I}_{\text{cont}}(\nu)} = \frac{{I}(\nu)}{{I}_{\text{cont}}(\nu)} - 1
\end{equation}
where $I(\nu)$ is the total intensity. Now, we can rewrite eq.~\ref{eq:eq6} in terms of $T_{\text{cont}}$, $T_{\text{line}}$, $\tau_{\text{cont}}(\nu)$ and $\tau_{\text{line}}(\nu)$ as follows,

\begin{equation}
\label{eq:eq7n}
    \frac{{T}_{\text{line}}}{{T}_{\text{cont}}} = \frac{ {e}^{-\tau_{cont}(\nu)} \left[1-{e}^{-\tau_{line}(\nu)}\right]}{\left[1-{e}^{-\tau_{cont}(\nu)}\right]}
\end{equation}

From eq.~\ref{eq:eq7n}, $\tau_{\text{line}}(\nu)$ can be expressed as

\begin{equation}
    \label{eq:eq8}
    \tau_{\text{line}}(\nu) = - \ln \left( 1-\frac{{T}_{\text{line}}}{{T}_{\text{cont}}}\, {e}^{\tau_{\text{cont}}(\nu)}\, \left[1-{e}^{-\tau_{\text{cont}}(\nu)}\right] \right)
\end{equation}

Using eq.~\ref{eq:eq3} and eq.~\ref{eq:eq8}, one can determine $\tau_{\text{cont}}(\nu)$ and $\tau_{\text{line}}(\nu)$ for each pixel. The value of $\tau_{\text{cont}}(\nu)$ along a line of sight (LOS) can also be related to the properties of the HII region using the Altenhoff approximation \citep{altenhoff1960veroffentl}.

\begin{equation}
\label{eq:eq9}
    \tau_{\text{cont}}(\nu) \approx 8.235 \times 10^{-2} \,\left(\frac{{T}_{\text{e}}}{\text{K}}\right)^{-1.35} \,\left(\frac{\nu}{\text{GHz}}\right)^{-2.1}\, \left(\frac{EM}{\text{pc}.\text{cm}^{-6}}\right)
\end{equation}
where $EM$ is the emission measure. Similarly, following \citet{Wilson2009}, the line optical depth is,

\begin{equation}
\label{eq:eq10}
    \tau_{\text{line}}(\nu) \approx 1.92 \times 10^{3} \,\left(\frac{{T}_{\text{e}}}{\text{K}}\right)^{-2.5} \,\left(\frac{\Delta\nu}{\text{KHz}}\right)^{-1}\, \left(\frac{EM}{\text{pc}.\text{cm}^{-6}}\right)
\end{equation}
for the recombination lines in the radio frequency domain. Here $\Delta\nu$ is the FWHM of the RRLs in KHz. Following eq.~\ref{eq:eq9} and eq.~\ref{eq:eq10} we get,

\begin{equation}
\label{eq:eq11}
    \frac{\tau_{\text{line}}(\nu)}{\tau_{\text{cont}}(\nu)} = 23.315 \times 10^{3} \,\left(\frac{{T}_{\text{e}}}{\text{K}}\right)^{-1.15} \,\left(\frac{\Delta\nu}{\text{KHz}}\right)^{-1}\, \left(\frac{\nu}{GHz}\right)^{2.1}
\end{equation}

Since the optical depth of the line and continuum are measured from the data, we can use eq.~\ref{eq:eq11} to estimate pixel-to-pixel $T_{\text{e}}$ values for the entire region under the LTE conditions. It is to be noted that eq.~\ref{eq:eq3} is not completely independent. We have to provide a $T_{\text{e}}$ to derive a $\tau_{\text{cont}}(\nu)$ value. When the continuum is optically thin, the continuum optical depth is inversely proportional to $T_{\text{e}}$, and one can use eqs.~\ref{eq:eq7n} and~\ref{eq:eq11} to derive $T_{\text{e}}$. However, since the relation between the continuum optical depth and $T_{\text{e}}$ is non-linear for moderate optical depth, we adopt an iterative procedure, wherein an initial guess of the electron temperature is provided in eq.~\ref{eq:eq3} following which the electron temperature is re-computed using eq.~\ref{eq:eq11}. This is repeated until the value of electron temperature converges.

\citet{1966ApJ...144.1225G} showed that the intensities of the observed RRLs could only be explained using departures from LTE, similar to the anomalous intensities of optical lines from nebulae and stellar atmospheres. Following \citet{1966ApJ...144.1225G}, the excitation temperature ($T_{\text{ex}}$) of an electronic transition is not equal to the $T_{\text{e}}$ of the ionized gas inside a HII region, and is thus corrected using

\begin{equation}
    \label{eq:eq12}
    {e}^{-\,{h}\,\nu\,/\,{k}_{\text{B}} {T}_{\text{ex}}} = 
    \frac{{b}_{\text{n}}}{{b}_{\text{n-1}}}\,\,{e}^{-\,{h}\,\nu\,/\,{k}_{\text{B}} {T}_{\text{e}}}
\end{equation}
where $b_{\text{n}}$, the departure coefficient, is the ratio of the actual population of atoms in the n-th state to the population which would be there if the ionized gas was in the LTE at the temperature $T_{\text{e}}$. Under the condition $h\nu << k_{\text{B}}T_{\text{e}}$, the corrected or non-LTE value of the line absorption coefficient ($\kappa_{\text{line}}$) for the line n $\rightarrow$ n--1 is found to be \citep{1966ApJ...144.1225G, 1972MNRAS.157..179B},

\begin{equation}
    \label{eq:eq13}
    \kappa_{\text{line}} = \kappa^*_{\text{line}}\,{b}_{\text{n-1}}\,\beta
\end{equation}
where $\kappa^*_{\text{line}}$ is the LTE value of the line absorption coefficient, and the correction factor ($\beta$) is approximated by \citet{1972MNRAS.157..179B} for RRL transitions with $\Delta{n} \approx 1$. If the values of $|\tau_{\text{line}}|$ and $\tau_{\text{cont}}$ are less than 1, we can approximate the exponential terms in eq.~\ref{eq:eq7n} to the second order to get

\begin{equation}
    \label{eq:eq14}
    {T}_{\text{line}} \, \approx \, {T}^{*}_{\text{line}}\,{b}_{\text{n}} \, \left(1-\frac{\tau_{\text{cont}}}{2}\beta\right)
\end{equation}
under the condition $h\nu <<k_{\text{B}}T_{\text{e}}$. Again, ${T}^*_{\text{line}}$ is the LTE value of the line temperature. Now, using the non-LTE value i.e. ${T}_{\text{line}}$ in eq.~\ref{eq:eq8} and following eq.~\ref{eq:eq11}, we can get the non-LTE electron temperature of the ionized gas.


\bibliography{sample63}{}

\begin{thebibliography}{}
\expandafter\ifx\csname natexlab\endcsname\relax\def\natexlab#1{#1}\fi
\providecommand{\url}[1]{\href{#1}{#1}}
\providecommand{\dodoi}[1]{doi:~\href{http://doi.org/#1}{\nolinkurl{#1}}}
\providecommand{\doeprint}[1]{\href{http://ascl.net/#1}{\nolinkurl{http://ascl.net/#1}}}
\providecommand{\doarXiv}[1]{\href{https://arxiv.org/abs/#1}{\nolinkurl{https://arxiv.org/abs/#1}}}

\bibitem[{{Allen} {et~al.}(2004){Allen}, {Calvet}, {D'Alessio}, {Merin},
  {Hartmann}, {Megeath}, {Gutermuth}, {Muzerolle}, {Pipher}, {Myers}, \&
  {Fazio}}]{2004astro.ph..6003A}
{Allen}, L.~E., {Calvet}, N., {D'Alessio}, P., {et~al.} 2004, arXiv e-prints

\bibitem[{Altenhoff {et~al.}(1960)Altenhoff, Mezger, Wendker, \&
  Westerhout}]{altenhoff1960veroffentl}
Altenhoff, W., Mezger, P., Wendker, H., \& Westerhout, G. 1960, Sternwarte
  Bonn, 48

\bibitem[{{Aniano} {et~al.}(2011){Aniano}, {Draine}, {Gordon}, \&
  {Sandstrom}}]{2011PASP..123.1218A}
{Aniano}, G., {Draine}, B.~T., {Gordon}, K.~D., \& {Sandstrom}, K. 2011, \pasp,
  123, 1218

\bibitem[{{Baug} {et~al.}(2016){Baug}, {Dewangan}, {Ojha}, \&
  {Ninan}}]{2016ApJ...833...85B}
{Baug}, T., {Dewangan}, L.~K., {Ojha}, D.~K., \& {Ninan}, J.~P. 2016, \apj,
  833, 85

\bibitem[{{Benjamin} {et~al.}(2003){Benjamin}, {Churchwell}, {Babler}, {Bania},
  {Clemens}, {Cohen}, {Dickey}, {Indebetouw}, {Jackson}, {Kobulnicky},
  {Lazarian}, {Marston}, {Mathis}, {Meade}, {Seager}, {Stolovy}, {Watson},
  {Whitney}, {Wolff}, \& {Wolfire}}]{2003PASP..115..953B}
{Benjamin}, R.~A., {Churchwell}, E., {Babler}, B.~L., {et~al.} 2003, \pasp,
  115, 953

\bibitem[{{Bessell} \& {Brett}(1988)}]{1988PASP..100.1134B}
{Bessell}, M.~S., \& {Brett}, J.~M. 1988, \pasp, 100, 1134

\bibitem[{{Beuther} {et~al.}(2016){Beuther}, {Bihr}, {Rugel}, {Johnston},
  {Wang}, {Walter}, {Brunthaler}, {Walsh}, {Ott}, {Stil}, {Henning},
  {Schierhuber}, {Kainulainen}, {Heyer}, {Goldsmith}, {Anderson}, {Longmore},
  {Klessen}, {Glover}, {Urquhart}, {Plume}, {Ragan}, {Schneider},
  {McClure-Griffiths}, {Menten}, {Smith}, {Roy}, {Shanahan}, {Nguyen-Luong}, \&
  {Bigiel}}]{2016A&A...595A..32B}
{Beuther}, H., {Bihr}, S., {Rugel}, M., {et~al.} 2016, \aap, 595, A32

\bibitem[{{Bohlin} {et~al.}(1978){Bohlin}, {Savage}, \&
  {Drake}}]{1978ApJ...224..132B}
{Bohlin}, R.~C., {Savage}, B.~D., \& {Drake}, J.~F. 1978, \apj, 224, 132

\bibitem[{{Brocklehurst} \& {Seaton}(1972)}]{1972MNRAS.157..179B}
{Brocklehurst}, M., \& {Seaton}, M.~J. 1972, \mnras, 157, 179

\bibitem[{{Carpenter}(2001)}]{2001AJ....121.2851C}
{Carpenter}, J.~M. 2001, \aj, 121, 2851

\bibitem[{{Churchwell}(2002)}]{2002ARA&A..40...27C}
{Churchwell}, E. 2002, \araa, 40, 27

\bibitem[{{Condon} {et~al.}(1998){Condon}, {Cotton}, {Greisen}, {Yin},
  {Perley}, {Taylor}, \& {Broderick}}]{1998AJ....115.1693C}
{Condon}, J.~J., {Cotton}, W.~D., {Greisen}, E.~W., {et~al.} 1998, \aj, 115,
  1693

\bibitem[{{Corradi} {et~al.}(1998){Corradi}, {Aznar}, \&
  {Mampaso}}]{1998MNRAS.297..617C}
{Corradi}, R. L.~M., {Aznar}, R., \& {Mampaso}, A. 1998, \mnras, 297, 617

\bibitem[{{Dempsey} {et~al.}(2013){Dempsey}, {Thomas}, \&
  {Currie}}]{2013ApJS..209....8D}
{Dempsey}, J.~T., {Thomas}, H.~S., \& {Currie}, M.~J. 2013, \apjs, 209, 8

\bibitem[{{Dewangan} \& {Ojha}(2017)}]{2017ApJ...849...65D}
{Dewangan}, L.~K., \& {Ojha}, D.~K. 2017, \apj, 849, 65

\bibitem[{{Dewangan} {et~al.}(2018){Dewangan}, {Ojha}, {Zinchenko}, \&
  {Baug}}]{2018ApJ...861...19D}
{Dewangan}, L.~K., {Ojha}, D.~K., {Zinchenko}, I., \& {Baug}, T. 2018, \apj,
  861, 19

\bibitem[{{Dewangan} {et~al.}(2019){Dewangan}, {Sano}, {Enokiya}, {Tachihara},
  {Fukui}, \& {Ojha}}]{2019ApJ...878...26D}
{Dewangan}, L.~K., {Sano}, H., {Enokiya}, R., {et~al.} 2019, \apj, 878, 26

\bibitem[{{Dickel} {et~al.}(1978){Dickel}, {Dickel}, \&
  {Wilson}}]{1978ApJ...223..840D}
{Dickel}, J.~R., {Dickel}, H.~R., \& {Wilson}, W.~J. 1978, \apj, 223, 840

\bibitem[{{Dyson} \& {Williams}(1980)}]{1980pim..book.....D}
{Dyson}, J.~E., \& {Williams}, D.~A. 1980, {Physics of the interstellar medium}

\bibitem[{{Fukui} {et~al.}(2014){Fukui}, {Ohama}, {Hanaoka}, {Furukawa},
  {Torii}, {Dawson}, {Mizuno}, {Hasegawa}, {Fukuda}, {Soga}, {Moribe},
  {Kuroda}, {Hayakawa}, {Kawamura}, {Kuwahara}, {Yamamoto}, {Okuda}, {Onishi},
  {Maezawa}, \& {Mizuno}}]{2014ApJ...780...36F}
{Fukui}, Y., {Ohama}, A., {Hanaoka}, N., {et~al.} 2014, \apj, 780, 36

\bibitem[{{Furukawa} {et~al.}(2009){Furukawa}, {Dawson}, {Ohama}, {Kawamura},
  {Mizuno}, {Onishi}, \& {Fukui}}]{2009ApJ...696L.115F}
{Furukawa}, N., {Dawson}, J.~R., {Ohama}, A., {et~al.} 2009, \apjl, 696, L115

\bibitem[{{Goldberg}(1966)}]{1966ApJ...144.1225G}
{Goldberg}, L. 1966, \apj, 144, 1225

\bibitem[{{Gutermuth} {et~al.}(2008){Gutermuth}, {Myers}, {Megeath}, {Allen},
  {Pipher}, {Muzerolle}, {Porras}, {Winston}, \& {Fazio}}]{2008ApJ...674..336G}
{Gutermuth}, R.~A., {Myers}, P.~C., {Megeath}, S.~T., {et~al.} 2008, \apj, 674,
  336

\bibitem[{{G{\"u}ver} \& {{\"O}zel}(2009)}]{2009MNRAS.400.2050G}
{G{\"u}ver}, T., \& {{\"O}zel}, F. 2009, \mnras, 400, 2050

\bibitem[{{Habe} \& {Ohta}(1992)}]{1992PASJ...44..203H}
{Habe}, A., \& {Ohta}, K. 1992, \pasj, 44, 203

\bibitem[{{Hartmann} {et~al.}(2005){Hartmann}, {Megeath}, {Allen}, {Luhman},
  {Calvet}, {D'Alessio}, {Franco-Hernandez}, \& {Fazio}}]{2005ApJ...629..881H}
{Hartmann}, L., {Megeath}, S.~T., {Allen}, L., {et~al.} 2005, \apj, 629, 881

\bibitem[{{Haworth} {et~al.}(2015){Haworth}, {Tasker}, {Fukui}, {Torii},
  {Dale}, {Shima}, {Takahira}, {Habe}, \& {Hasegawa}}]{2015MNRAS.450...10H}
{Haworth}, T.~J., {Tasker}, E.~J., {Fukui}, Y., {et~al.} 2015, \mnras, 450, 10

\bibitem[{{Hoare} {et~al.}(2012){Hoare}, {Purcell}, {Churchwell}, {Diamond},
  {Cotton}, {Chandler}, {Smethurst}, {Kurtz}, {Mundy}, {Dougherty}, {Fender},
  {Fuller}, {Jackson}, {Garrington}, {Gledhill}, {Goldsmith}, {Lumsden},
  {Mart{\'\i}}, {Moore}, {Muxlow}, {Oudmaijer}, {Pandian}, {Paredes},
  {Shepherd}, {Spencer}, {Thompson}, {Umana}, {Urquhart}, \&
  {Zijlstra}}]{2012PASP..124..939H}
{Hoare}, M.~G., {Purcell}, C.~R., {Churchwell}, E.~B., {et~al.} 2012, \pasp,
  124, 939

\bibitem[{{Inoue} \& {Fukui}(2013)}]{2013ApJ...774L..31I}
{Inoue}, T., \& {Fukui}, Y. 2013, \apjl, 774, L31

\bibitem[{{Issac} {et~al.}(2020){Issac}, {Tej}, {Liu}, \&
  {Wu}}]{2020MNRAS.499.3620I}
{Issac}, N., {Tej}, A., {Liu}, T., \& {Wu}, Y. 2020, \mnras, 499, 3620

\bibitem[{{Jolly} {et~al.}(2020){Jolly}, {Knudsen}, \&
  {Stanley}}]{2020MNRAS.499.3992J}
{Jolly}, J.-B., {Knudsen}, K.~K., \& {Stanley}, F. 2020, \mnras, 499, 3992

\bibitem[{{Kratter} \& {Matzner}(2006)}]{2006MNRAS.373.1563K}
{Kratter}, K.~M., \& {Matzner}, C.~D. 2006, \mnras, 373, 1563

\bibitem[{{Krumholz} {et~al.}(2009){Krumholz}, {Klein}, {McKee}, {Offner}, \&
  {Cunningham}}]{2009Sci...323..754K}
{Krumholz}, M.~R., {Klein}, R.~I., {McKee}, C.~F., {Offner}, S. S.~R., \&
  {Cunningham}, A.~J. 2009, Science, 323, 754

\bibitem[{{Kurtz} {et~al.}(1994){Kurtz}, {Churchwell}, \&
  {Wood}}]{1994ApJS...91..659K}
{Kurtz}, S., {Churchwell}, E., \& {Wood}, D.~O.~S. 1994, \apjs, 91, 659

\bibitem[{{Lawrence} {et~al.}(2007){Lawrence}, {Warren}, {Almaini}, {Edge},
  {Hambly}, {Jameson}, {Lucas}, {Casali}, {Adamson}, {Dye}, {Emerson},
  {Foucaud}, {Hewett}, {Hirst}, {Hodgkin}, {Irwin}, {Lodieu}, {McMahon},
  {Simpson}, {Smail}, {Mortlock}, \& {Folger}}]{2007MNRAS.379.1599L}
{Lawrence}, A., {Warren}, S.~J., {Almaini}, O., {et~al.} 2007, \mnras, 379,
  1599

\bibitem[{{Li} \& {Klein}(2019)}]{2019MNRAS.485.4509L}
{Li}, P.~S., \& {Klein}, R.~I. 2019, \mnras, 485, 4509

\bibitem[{{Lockman}(1989)}]{1989ApJS...71..469L}
{Lockman}, F.~J. 1989, \apjs, 71, 469

\bibitem[{{Loren}(1976)}]{1976ApJ...209..466L}
{Loren}, R.~B. 1976, \apj, 209, 466

\bibitem[{{Loren}(1977)}]{1977ApJ...215..129L}
---. 1977, \apj, 215, 129

\bibitem[{{Martins} {et~al.}(2005){Martins}, {Schaerer}, \&
  {Hillier}}]{2005A&A...436.1049M}
{Martins}, F., {Schaerer}, D., \& {Hillier}, D.~J. 2005, \aap, 436, 1049

\bibitem[{{McKee} \& {Tan}(2003)}]{2003ApJ...585..850M}
{McKee}, C.~F., \& {Tan}, J.~C. 2003, \apj, 585, 850

\bibitem[{{McMullin} {et~al.}(2007){McMullin}, {Waters}, {Schiebel}, {Young},
  \& {Golap}}]{2007ASPC..376..127M}
{McMullin}, J.~P., {Waters}, B., {Schiebel}, D., {Young}, W., \& {Golap}, K.
  2007, in Astronomical Society of the Pacific Conference Series, Vol. 376,
  Astronomical Data Analysis Software and Systems XVI, 127

\bibitem[{{Meyer} {et~al.}(1997){Meyer}, {Calvet}, \&
  {Hillenbrand}}]{1997AJ....114..288M}
{Meyer}, M.~R., {Calvet}, N., \& {Hillenbrand}, L.~A. 1997, \aj, 114, 288

\bibitem[{{Mezger} \& {Henderson}(1967)}]{Mezger67}
{Mezger}, P.~G., \& {Henderson}, A.~P. 1967, \apj, 147, 471

\bibitem[{{Molinari} {et~al.}(2010){Molinari}, {Swinyard}, {Bally}, {Barlow},
  {Bernard}, {Martin}, {Moore}, {Noriega-Crespo}, {Plume}, {Testi}, {Zavagno},
  {Abergel}, {Ali}, {Andr{\'e}}, {Baluteau}, {Benedettini}, {Bern{\'e}},
  {Billot}, {Blommaert}, {Bontemps}, {Boulanger}, {Brand}, {Brunt}, {Burton},
  {Campeggio}, {Carey}, {Caselli}, {Cesaroni}, {Cernicharo}, {Chakrabarti},
  {Chrysostomou}, {Codella}, {Cohen}, {Compiegne}, {Davis}, {de Bernardis}, {de
  Gasperis}, {Di Francesco}, {di Giorgio}, {Elia}, {Faustini}, {Fischera},
  {Fukui}, {Fuller}, {Ganga}, {Garcia-Lario}, {Giard}, {Giardino}, {Glenn},
  {Goldsmith}, {Griffin}, {Hoare}, {Huang}, {Jiang}, {Joblin}, {Joncas},
  {Juvela}, {Kirk}, {Lagache}, {Li}, {Lim}, {Lord}, {Lucas}, {Maiolo},
  {Marengo}, {Marshall}, {Masi}, {Massi}, {Matsuura}, {Meny}, {Minier},
  {Miville-Desch{\^e}nes}, {Montier}, {Motte}, {M{\"u}ller}, {Natoli}, {Neves},
  {Olmi}, {Paladini}, {Paradis}, {Pestalozzi}, {Pezzuto}, {Piacentini},
  {Pomar{\`e}s}, {Popescu}, {Reach}, {Richer}, {Ristorcelli}, {Roy}, {Royer},
  {Russeil}, {Saraceno}, {Sauvage}, {Schilke}, {Schneider-Bontemps},
  {Schuller}, {Schultz}, {Shepherd}, {Sibthorpe}, {Smith}, {Smith},
  {Spinoglio}, {Stamatellos}, {Strafella}, {Stringfellow}, {Sturm}, {Taylor},
  {Thompson}, {Tuffs}, {Umana}, {Valenziano}, {Vavrek}, {Viti}, {Waelkens},
  {Ward-Thompson}, {White}, {Wyrowski}, {Yorke}, \&
  {Zhang}}]{2010PASP..122..314M}
{Molinari}, S., {Swinyard}, B., {Bally}, J., {et~al.} 2010, \pasp, 122, 314

\bibitem[{{Ohama} {et~al.}(2010){Ohama}, {Dawson}, {Furukawa}, {Kawamura},
  {Moribe}, {Yamamoto}, {Okuda}, {Mizuno}, {Onishi}, {Maezawa}, {Minamidani},
  {Mizuno}, \& {Fukui}}]{2010ApJ...709..975O}
{Ohama}, A., {Dawson}, J.~R., {Furukawa}, N., {et~al.} 2010, \apj, 709, 975

\bibitem[{{Ossenkopf} \& {Henning}(1994)}]{1994A&A...291..943O}
{Ossenkopf}, V., \& {Henning}, T. 1994, \aap, 291, 943

\bibitem[{{Osterbrock}(1989)}]{1989agna.book.....O}
{Osterbrock}, D.~E. 1989, {Astrophysics of gaseous nebulae and active galactic
  nuclei}

\bibitem[{{Peretto} \& {Fuller}(2009)}]{2009A&A...505..405P}
{Peretto}, N., \& {Fuller}, G.~A. 2009, \aap, 505, 405

\bibitem[{{Planck Collaboration} {et~al.}(2015){Planck Collaboration}, {Ade},
  {Aghanim}, {Alina}, {Alves}, {Armitage-Caplan}, {Arnaud}, {Arzoumanian},
  {Ashdown}, {Atrio-Barandela}, {Aumont}, {Baccigalupi}, {Banday}, {Barreiro},
  {Battaner}, {Benabed}, {Benoit-L{\'e}vy}, {Bernard}, {Bersanelli},
  {Bielewicz}, {Bock}, {Bond}, {Borrill}, {Bouchet}, {Boulanger}, {Bracco},
  {Burigana}, {Butler}, {Cardoso}, {Catalano}, {Chamballu}, {Chary}, {Chiang},
  {Christensen}, {Colombi}, {Colombo}, {Combet}, {Couchot}, {Coulais}, {Crill},
  {Curto}, {Cuttaia}, {Danese}, {Davies}, {Davis}, {de Bernardis}, {de Gouveia
  Dal Pino}, {de Rosa}, {de Zotti}, {Delabrouille}, {D{\'e}sert}, {Dickinson},
  {Diego}, {Donzelli}, {Dor{\'e}}, {Douspis}, {Dunkley}, {Dupac}, {Efstathiou},
  {En{\ss}lin}, {Eriksen}, {Falgarone}, {Ferri{\`e}re}, {Finelli}, {Forni},
  {Frailis}, {Fraisse}, {Franceschi}, {Galeotta}, {Ganga}, {Ghosh}, {Giard},
  {Giraud-H{\'e}raud}, {Gonz{\'a}lez-Nuevo}, {G{\'o}rski}, {Gregorio},
  {Gruppuso}, {Guillet}, {Hansen}, {Harrison}, {Helou},
  {Hern{\'a}ndez-Monteagudo}, {Hildebrandt}, {Hivon}, {Hobson}, {Holmes},
  {Hornstrup}, {Huffenberger}, {Jaffe}, {Jaffe}, {Jones}, {Juvela},
  {Keih{\"a}nen}, {Keskitalo}, {Kisner}, {Kneissl}, {Knoche}, {Kunz},
  {Kurki-Suonio}, {Lagache}, {L{\"a}hteenm{\"a}ki}, {Lamarre}, {Lasenby},
  {Lawrence}, {Leahy}, {Leonardi}, {Levrier}, {Liguori}, {Lilje},
  {Linden-V{\o}rnle}, {L{\'o}pez-Caniego}, {Lubin}, {Mac{\'\i}as-P{\'e}rez},
  {Maffei}, {Magalh{\~a}es}, {Maino}, {Mandolesi}, {Maris}, {Marshall},
  {Martin}, {Mart{\'\i}nez-Gonz{\'a}lez}, {Masi}, {Matarrese}, {Mazzotta},
  {Melchiorri}, {Mendes}, {Mennella}, {Migliaccio}, {Miville-Desch{\^e}nes},
  {Moneti}, {Montier}, {Morgante}, {Mortlock}, {Munshi}, {Murphy}, {Naselsky},
  {Nati}, {Natoli}, {Netterfield}, {Noviello}, {Novikov}, {Novikov},
  {Oxborrow}, {Pagano}, {Pajot}, {Paladini}, {Paoletti}, {Pasian}, {Pearson},
  {Perdereau}, {Perotto}, {Perrotta}, {Piacentini}, {Piat}, {Pietrobon},
  {Plaszczynski}, {Poidevin}, {Pointecouteau}, {Polenta}, {Popa}, {Pratt},
  {Prunet}, {Puget}, {Rachen}, {Reach}, {Rebolo}, {Reinecke}, {Remazeilles},
  {Renault}, {Ricciardi}, {Riller}, {Ristorcelli}, {Rocha}, {Rosset},
  {Roudier}, {Rubi{\~n}o-Mart{\'\i}n}, {Rusholme}, {Sandri}, {Savini}, {Scott},
  {Spencer}, {Stolyarov}, {Stompor}, {Sudiwala}, {Sutton}, {Suur-Uski},
  {Sygnet}, {Tauber}, {Terenzi}, {Toffolatti}, {Tomasi}, {Tristram}, {Tucci},
  {Umana}, {Valenziano}, {Valiviita}, {Van Tent}, {Vielva}, {Villa}, {Wade},
  {Wandelt}, {Zacchei}, \& {Zonca}}]{2015A&A...576A.104P}
{Planck Collaboration}, {Ade}, P.~A.~R., {Aghanim}, N., {et~al.} 2015, \aap,
  576, A104

\bibitem[{{Predehl} \& {Schmitt}(1995)}]{1995A&A...293..889P}
{Predehl}, P., \& {Schmitt}, J.~H.~M.~M. 1995, \aap, 500, 459

\bibitem[{{Priestley} \& {Whitworth}(2021)}]{2021MNRAS.506..775P}
{Priestley}, F.~D., \& {Whitworth}, A.~P. 2021, \mnras, 506, 775

\bibitem[{{Purcell} {et~al.}(2013){Purcell}, {Hoare}, {Cotton}, {Lumsden},
  {Urquhart}, {Chandler}, {Churchwell}, {Diamond}, {Dougherty}, {Fender},
  {Fuller}, {Garrington}, {Gledhill}, {Goldsmith}, {Hindson}, {Jackson},
  {Kurtz}, {Mart{\'\i}}, {Moore}, {Mundy}, {Muxlow}, {Oudmaijer}, {Pandian},
  {Paredes}, {Shepherd}, {Smethurst}, {Spencer}, {Thompson}, {Umana}, \&
  {Zijlstra}}]{2013ApJS..205....1P}
{Purcell}, C.~R., {Hoare}, M.~G., {Cotton}, W.~D., {et~al.} 2013, \apjs, 205, 1

\bibitem[{{Quireza} {et~al.}(2006){Quireza}, {Rood}, {Bania}, {Balser}, \&
  {Maciel}}]{2006ApJ...653.1226Q}
{Quireza}, C., {Rood}, R.~T., {Bania}, T.~M., {Balser}, D.~S., \& {Maciel},
  W.~J. 2006, \apj, 653, 1226

\bibitem[{{Rieke} \& {Lebofsky}(1985)}]{1985ApJ...288..618R}
{Rieke}, G.~H., \& {Lebofsky}, M.~J. 1985, \apj, 288, 618

\bibitem[{{Russeil} {et~al.}(2013){Russeil}, {Schneider}, {Anderson},
  {Zavagno}, {Molinari}, {Persi}, {Bontemps}, {Motte}, {Ossenkopf},
  {Andr{\'e}}, {Arzoumanian}, {Bernard}, {Deharveng}, {Didelon}, {Di
  Francesco}, {Elia}, {Hennemann}, {Hill}, {K{\"o}nyves}, {Li}, {Martin},
  {Nguyen Luong}, {Peretto}, {Pezzuto}, {Polychroni}, {Roussel}, {Rygl},
  {Spinoglio}, {Testi}, {Tig{\'e}}, {Vavrek}, {Ward-Thompson}, \&
  {White}}]{2013A&A...554A..42R}
{Russeil}, D., {Schneider}, N., {Anderson}, L.~D., {et~al.} 2013, \aap, 554,
  A42

\bibitem[{{Schmiedeke} {et~al.}(2016){Schmiedeke}, {Schilke}, {M{\"o}ller},
  {S{\'a}nchez-Monge}, {Bergin}, {Comito}, {Csengeri}, {Lis}, {Molinari},
  {Qin}, \& {Rolffs}}]{2016A&A...588A.143S}
{Schmiedeke}, A., {Schilke}, P., {M{\"o}ller}, T., {et~al.} 2016, \aap, 588,
  A143

\bibitem[{{Simon} {et~al.}(2006){Simon}, {Jackson}, {Rathborne}, \&
  {Chambers}}]{2006ApJ...639..227S}
{Simon}, R., {Jackson}, J.~M., {Rathborne}, J.~M., \& {Chambers}, E.~T. 2006,
  \apj, 639, 227

\bibitem[{{Skrutskie} {et~al.}(2006){Skrutskie}, {Cutri}, {Stiening},
  {Weinberg}, {Schneider}, {Carpenter}, {Beichman}, {Capps}, {Chester},
  {Elias}, {Huchra}, {Liebert}, {Lonsdale}, {Monet}, {Price}, {Seitzer},
  {Jarrett}, {Kirkpatrick}, {Gizis}, {Howard}, {Evans}, {Fowler}, {Fullmer},
  {Hurt}, {Light}, {Kopan}, {Marsh}, {McCallon}, {Tam}, {Van Dyk}, \&
  {Wheelock}}]{2005sptz.prop20710S}
{Skrutskie}, M.~F., {Cutri}, R.~M., {Stiening}, R., {et~al.} 2006, \aj, 131,
  1163

\bibitem[{{Swarup}(1990)}]{1990IJRSP..19..493S}
{Swarup}, G. 1990, Indian Journal of Radio and Space Physics, 19, 493

\bibitem[{{Takahira} {et~al.}(2018){Takahira}, {Shima}, {Habe}, \&
  {Tasker}}]{2018PASJ...70S..58T}
{Takahira}, K., {Shima}, K., {Habe}, A., \& {Tasker}, E.~J. 2018, \pasj, 70,
  S58

\bibitem[{{Torii} {et~al.}(2015){Torii}, {Hasegawa}, {Hattori}, {Sano},
  {Ohama}, {Yamamoto}, {Tachihara}, {Soga}, {Shimizu}, {Okuda}, {Mizuno},
  {Onishi}, {Mizuno}, \& {Fukui}}]{2015ApJ...806....7T}
{Torii}, K., {Hasegawa}, K., {Hattori}, Y., {et~al.} 2015, \apj, 806, 7

\bibitem[{{Urquhart} {et~al.}(2011){Urquhart}, {Morgan}, {Figura}, {Moore},
  {Lumsden}, {Hoare}, {Oudmaijer}, {Mottram}, {Davies}, \&
  {Dunham}}]{2011MNRAS.418.1689U}
{Urquhart}, J.~S., {Morgan}, L.~K., {Figura}, C.~C., {et~al.} 2011, \mnras,
  418, 1689

\bibitem[{{Ward-Thompson} {et~al.}(2010){Ward-Thompson}, {Kirk}, {Andr{\'e}},
  {Saraceno}, {Didelon}, {K{\"o}nyves}, {Schneider}, {Abergel}, {Baluteau},
  {Bernard}, {Bontemps}, {Cambr{\'e}sy}, {Cox}, {di Francesco}, {di Giorgio},
  {Griffin}, {Hargrave}, {Huang}, {Li}, {Martin}, {Men'shchikov}, {Minier},
  {Molinari}, {Motte}, {Olofsson}, {Pezzuto}, {Russeil}, {Sauvage},
  {Sibthorpe}, {Spinoglio}, {Testi}, {White}, {Wilson}, {Woodcraft}, \&
  {Zavagno}}]{2010A&A...518L..92W}
{Ward-Thompson}, D., {Kirk}, J.~M., {Andr{\'e}}, P., {et~al.} 2010, \aap, 518,
  L92

\bibitem[{Wilson {et~al.}(2009)Wilson, Rohlfs, \&
  H{\"u}ttemeister}]{Wilson2009}
Wilson, T.~L., Rohlfs, K., \& H{\"u}ttemeister, S. 2009, 319--328

\bibitem[{{Wolfire} \& {Cassinelli}(1987)}]{1987ApJ...319..850W}
{Wolfire}, M.~G., \& {Cassinelli}, J.~P. 1987, \apj, 319, 850

\bibitem[{{Zhang} {et~al.}(2017){Zhang}, {Yuan}, {Xu}, {Liu}, {Yu}, {Li}, {He},
  {Zhang}, \& {Wang}}]{2017RAA....17...57Z}
{Zhang}, C.-P., {Yuan}, J.-H., {Xu}, J.-L., {et~al.} 2017, Research in
  Astronomy and Astrophysics, 17, 057

\end{thebibliography}
\bibliographystyle{aasjournal}



\end{document}